# PORTABLE DEVICES FOR IN VITRO CHARACTERIZATION BASED ON ULTRASOUND

A Thesis

by

Seyedfakhreddin Nabavi

Submitted to the
Graduate School of Sciences and Engineering
In Partial Fulfillment of the Requirements for
the Degree of

Master of Science

in the
Department of Electrical and Electronics Engineering

Özyeğin University
August 2015

# PORTABLE DEVICES FOR IN VITRO CHARACTERIZATION BASED ON ULTRASOUND

Approved by:

---

Professor Goksen G. Yaralioglu, Advisor
Department of Electrical and Electronics Engineering
*Özyeğin University*

---

Professor Fatih Uğurdağ
Department of Electrical and Electronics Engineering
*Özyeğin University*

---

Professor Ayhan Bozkurt
Department of Electrical and Electronics Engineering
*Sabanci University*

Date Approved: 19 August 2015

*To My Parents...*

# ABSTRACT

We propose a portable cartridge based lab-on-a-chip platform. The main sensing mechanism uses high frequency acoustic waves. Feasibility of two different transducers has been investigated for biological measurements. We used zincoxide (ZnO) based piezoelectric transducers for blood coagulation time measurements. The method is based on the measurement of amplitude of acoustic reflections from a microfluidic channel which is filled with blood. The volume of the microfluidic channel could be as low as 1 micro-litre. A low cost disposable cartridge made of glass has been designed to perform the measurements. The cartridge is composed of two glass substrates and a double sided tape layer where the channel is defined. One of the glass substrates has a micromachined ZnO based transducer that operates at 400 MHz on the outer surface of the cartridge where electrical connections are also provided. The transducer is aligned with the channel to generate acoustic pulses in the fluid that is filling the channel. The reflections coming from the top of the channel propagate through the liquid therefore their amplitude and phase are affected by the fluid properties. In the experiments whole blood was used without any sample preparation. The viscosity of the blood changes during coagulation therefore by monitoring the amplitude of the reflection coming from the top of the channel one can measure the coagulation time. The method which requires only 1 micro-litre of blood has been tested with calcium and activated partial thromboplastin (aPTT) reagents. The proposed method has a potential to be used in a low cost portable coagulation time measurement system for patient self-testing. The second transducer was made of CMUTs (Capacitive Micromachined Ultrasonic Transducers). We introduce this technology for immersion in liquids as a biological sensor for viscosity measurements. Boundary conditions for membrane type resonators have been modified in order to increase the quality factor. This enabled a sensor with high sensitivity. We demonstrate the quality factor improvement using FEM analysis.

# ÖZETÇE

Bu projede taşınabilir kartuş tabanlı lab on a chip platformu öneriliyor. Ve bu platformun ana algılama mekanizması yüksek frekanslı akustik dalgalar ile çalışmaktadır. Biyolojik ölçümler için iki farklı transdüserin uygulanabilirliği incelenmiştir. Kan pıhtılaşma zamanı ölçümleri için ise çinkooksit (ZnO) tabanlı piezoelektrik transdüserler kullanılmıştır. Yöntem olarak kan ile dolu olan bir mikroakışkan kanaldan akustik yansımaların genliğinin ölçülmesi kullanımıştır. Mikroakışkan kanal hacmi 1 mikrolitre gibi düşük bir değer olabilir. Düşük maliyetli tek kullanımlık cam kartuş ise ölçümlerin yapılması için tasarlanmıştır. Kartuş iki cam alt tabakadan ve kanalın bulunduğu bir çift taraflı şerit tabakadan oluşur. Cam alt tabakalarından biri kartuşun dış yüzeyi üzerinde elektrik bağlantıları da sağlanmışbir şekilde 400 MHz'de çalışan bir mikroişlenmiş ZnO transdüsere sahiptir. Transdüser ise kanalı dolduran sıvı içinde akustik darbeleri üretmek için kanal ile hizalanmıştır. Kanalın üstünden gelen yansımalar sıvı içinden yayılmakta ve bu nedenle onların genlik ve fazları sıvının özelliklerinden dolayı etkilemektedir. Deneylerde tüm kan herhangi bir numune hazırlanmadan kullanılmıştır. Pıhtılaşma sırasında kan akışmazlı değişir dolayısıyla kanalın üst kısmından gelen yansıma genlii izlenerek pıhtılaşma süresi ölçülebilir. Kanın sadece 1 mikro-litre gerektiren yöntem kalsiyum ile test edilmiş ve parsiyel tromboplastin (aPTT) reaktifler ile aktive edilmiştir. önerilen yöntem, hastanın kendisini test etmek için kullanabileceği pıhtılaşma zamanını ölçen düşük maliyetli ve taşınabilir bir sistem için potansiyele sahiptir. Ikinci transdüser CMUTs olarak yapılmıştır. Bu teknolojiyi akışmazlık ölçümleri için sıvıların iine batırılan bir biyolojik sensr olarak tanıtıyoruz. Membran tipi rezonatörler için sınır koşulları kalite faktörünü arttırmak amacıyla değiştirilmiştir. Bu değişim sensöre yüksek hassasiyet kazandırmıştır. Kalite faktöründeki iyileşme sonlu eleman analizi kullanarak gösterilmiştir.

# ACKNOWLEDGEMENTS

First and foremost, I would like to thank my advisor professor Goksen G. Yaralioglu. He provided the golden opportunity of doing several interesting and attractive researches under his advice. He made this work possible and brilliant by giving the academic advices which are crucial for a student. It has been a pleasure for me to work under his impressive supervision and I will never forget his support. Throughout my education in Turkey, professor Yaralioglu helped me grow both professionally and personally. Once more time, I'm grateful for his favour.

I deeply thank my thesis committee members Dr. Ugurdag and Dr. Bozturkt for their interest in my work and their valuable guidances.

I also wish to offer my gratitude to all the colleagues specially my dear friend Mohammad Rahim Sobhani at the MEMS/Nano laboratory in the School of Engineering of Ozyegin university.

Last but not least, I want to thank my parents Mahmoud and Behi for their love and support and my dear brother Kaveh for his everlasting guidance and encouragement.

# TABLE OF CONTENTS

# LIST OF TABLES

# LIST OF FIGURES

# CHAPTER I

# INTRODUCTION

The small amplitude waves which are able to propagate in any environment such as gas, liquids and solids call acoustic waves. Acoustic waves carry energy and due to the attenuation factor in different materials the energy decay after passing a certain time.

The acoustic waves speed depends on material properties it propagates through. Basically, in materials with high density the speed of acoustic waves is higher than in materials with low density. For instance, in glass at the room temperature sound velocity is 5600 m/sec and in blood is 1500 m/sec. Also temperature affects on both attenuation and sound velocity.

The acoustic waves with the frequencies above 20 kHz call ultrasound waves which they are not detectable by human hearing system. These ultrasonic waves can be utilized in many areas such as industrial measurements, medical imaging and materials properties detection. Therefore, ultrasonic transducers and electronic devises are developing for producing the high frequency signals since material properties characterization in different environments hence many researches have be done on this area in past decades.

The piezoelectric materials are widely using for producing the ultrasonic waves. The transducers based on piezoelectric are able to generate the high frequency waves by applying the excitation signal to transducers. Another method for ultrasonic waves producing is using CMUT technology.

Capacitive Micromachined Ultrasonic Transducers (CMUT), using micromachined technique on silicon substrate a cavity build and a thin layer on top of this cavity located. Top and bottom electrodes on up and down of the cavity adjacent. If AC voltage applied to electrodes the membrane vibrated and ultrasonic waves will generate. This ultrasonic waves consist of high frequency and resolution for ultrasonic applications specially ultrasonic imaging.

## 1.1  *Thesis Goal*

This thesis presents a low coast cartridge that integrated with Zinc Oxide (ZnO) ultrasonic transducers based on piezoelectric for biological detection. The proposed sensor is completely portable and experimental results took place on whole blood for blood coagulation time measurement. Results show the proposed method has a potential to be used in a portable coagulation time measurement system for patient self-testing.

Furthermore, current resonant mass sensors have low Q factor where operate in liquids. So high quality factor resonant mass sensor for immersion in liquids based on CMUT technology in this thesis introduced. This sensor is able to vibrate inside the liquids environment with high quality factor. The high Q factor provides the high sensitivity hence it's advantage of this sensor. The method that described in this thesis could solve low quality factor problem of resonant sensors in liquids. Thus, featured small size and low cost will enable a portable sensing platform for biological and chemical tests.

# CHAPTER II

# ULTRASONIC TRANSDUCERS

## *2.1 Introduction*

The ultrasonic transducers can transmit and receive the ultrasonic waves. In this thesis the piezoelectric transducers (Zinc Oxide) transmitted and received wave by itself. Due to portable device purpose in this thesis, the required ultrasonic transducers for doing the experiments has been fabricated in 500 µm by 500 µm aspects and centre frequency for this transducer is 400 MHz. In following section further details about fabrication of this type transducers will discuss.

In additional, Capacitive Micromachined Ultrasonic Transducers (CMUTs) for using as a mechanical resonator introduced. This type of transducers also can operate as a transmitter and receiver of ultrasonic waves. CMUT technology is able to provide the high resonant frequency and high Q factor then, amount of the sensitivity in CMUTs transducers is high. Consequently, Capacitive Micromachined Ultrasonic Transducers (CMUTs) is a good option for resonant mass sensors in chemical and biological applications.

## *2.2 Zinc Oxide transducers*

Zinc Oxide is a direct semiconductor with wurtzite structure. The minimum energy gap is 3.2 eV at room temperature and 3.44 eV at 4 K. Some important members of wurtzite structure family such as ZnO, GaN, AlN, ZnS, and CdSe, which are important materials for applications in opto-electronics, lasing, and piezoelectricity. The two important characteristics of the wurtzite structure are the non-central symmetry and polar surfaces. The structure of ZnO illustrated in figure 1.

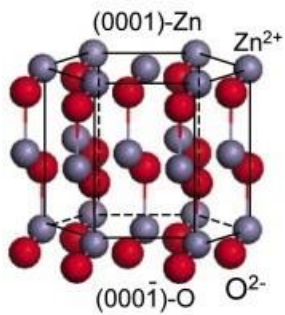

**Figure 1:** Prototypical ZnO schematic drawing. Figure reprinted from [1].

Due to ZnO material properties, it is a unique candidate for utilizing in many applications. For instance, ZnO widely used in mechanical, electrical, chemical, and optical areas. In some applications such as transparent conducting for displaying purpose and solar energy cells, surface and bulk acoustic wave devices (SAW), acoustic-optical devices, light emitting diodes (LEDs) and laser diodes (LDs), zinc oxide plays important role. Another Zinc Oxide privilege in compared to other materials is low cost, hence it has a potential option for industrial applications [1].

Numerous technologies have been used for manufacturing of ZnO thin films, containing chemical vapour deposition, solgel, spray-pyrolysis, molecular beam epitaxy, pulsed laser deposition, vacuum arc deposition and magnetron sputtering.

Compared to other thin-film deposition methods, magnetron sputtering technique has several advantages:

1. The deposition rate is high [2].

2. The smooth thickness with high density.

3. All material with different pressure can be sputtered easily.

4. Good stability and controllable condition during the process.

5. good adhesion of films on substrates.

6. The inexpensive deposition [3].

In 1936 first magnetron sputtering reported by Penning, until the planar magnetron invented, which is the suitable in sputtering deposition today. Basically, the feature of a magnetron discharge is used. This is achieved by the combination of electric and magnetic fields. By changing the magnetic field strength, electrons can be attached target while ions are not exist. The electrons orbits in interaction with magnetic and electric fields intend to ionization. Thus, magnetron discharges occur in low pressure and/or high current densities. More details about magnetron sputtering can be found in [4] and [3].

Our transducers based on Zinc Oxide in this thesis fabricated by magnetron sputtering method on glass surface and more information about the transducer aspects will discuss in next section.

## 2.3  *Capacitive Micromachined Ultrasonic Transducers (CMUT)*

Capacitive Micromachined Ultrasonic Transducer (CMUT) is a new principle in the ultrasonic transducers field. Generally, in CMUTs technology due to conversion in capacitance the sufficient energy for ultrasonic wave generated. CMUTs are manufactured on silicon substrate by utilizing the micromachining technique. A cavity is patterned in a silicon substrate, and a thin layer surrounded top of the cavity as a membrane. This thin layer is moveable member. On silicon substrate is bottom electrode located. If the AC signal is applied to biased electrodes (top and bottom electrode) the membrane vibrated due to capacitance change, then ultrasonic waves produced in the environment of interest. Thus, it works as a transmitter. Additionally, if ultrasonic waves are applied on the membranes of biased CMUT, it able to generate periodic signal due to changing the capacitance of the CMUT cell. In other words, under this condition varied capacitance appears. Thus, it works as a receiver by interaction ultrasonic waves with membrane. Figure 2 is a cross section view of single membrane for the CMUT transducer [5].

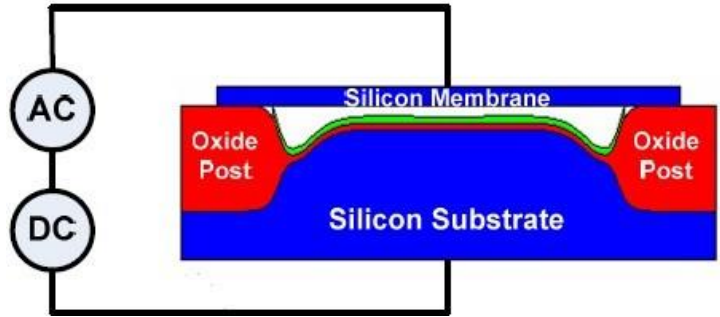

**Figure 2:** Schematic drawing of a single cell of a CMUT. Figure reprinted from [5].

CMUT cells are fabricated by using the micro-machined technology, it capable to manufacture 2D and 3D arrays of transducers, which means numerous of CMUTs could be included in a transducer array. The large number of CMUTs array is supplying larger bandwidth compared to other transducer technologies. Furthermore, due to small dimension of CMUTs transducers, it can operate with high frequency. The centre operation frequency and quality factor extremely in these type of transducers depend on the cell size (membranes), and on the materials stiffness that used in membranes fabrication. Another advantage of CMUTs transducers is they can be integrated electronics easily which other transducer technology types are not able providing this situation. According to CMUTs properties, high quality factor (in air) and high resonant frequency, make it appropriate option for using as a ultrasonic transducer for biological detection.

Resonant mass sensors can detect nano-sized particles to individual molecules without any labelling. Labelling increases the cost and complexity of the tests therefore it is not desirable. In biological applications, the structure is vibrated in aqueous solutions. Resonant mass sensors detect small changes in the resonant frequency of the mechanical structure due to the additional mass of the target particle that binds to the sensor surface.

In this thesis, new geometry for resonant mass sensor based on CMUTs transducers reported for immersion in liquids. As noted previously, the CMUTs technologies have high resonant frequency and high quality factor in air, however, the quality factor for CMUTs drops where it vibrates in the liquids medium. Therefore, the introduced new geometry could overcome this problem. In chapter 'Resonant mass sensor based on CMUT' the proposed sensor with further details will describe.

# CHAPTER III

# BLOOD COAGULATION TIME MEASUREMENT

## *3.1 Introduction*

Blood is an inhomogeneous tissue and blood coagulation is a dynamic physiological process which occurs with the activation of a series of enzyme for eliminating haemorrhage. There are various factors, enzymes and proteins that affect the clotting process. Thrombocytes also called Platelets have fundamental task to stop bleeding after an injury. Generally, the process of blood coagulation takes place in three crucial steps: (1) in reaction to rupture of the vessels or abruptly devastation in the blood; (2) the prothrombin activator from previous step accelerates the reduction prothrombin into thrombin; (3) the thrombin operates like an enzyme to convert fibrinogen (factor I) into fibrin (factor Ia). Eventually, after these complex cascades of chemical reactions the blood formation from liquid state transformed into solid gel. Thus, this in vivo actuality called blood coagulation [6].

Some drugs such as heparin or warfarin, an activated blood coagulation is used as the anticoagulant. Monitoring the blood coagulation for these patients is extremely important because the amount of anticoagulant drugs administered is determined from the coagulation time. In other words, when the anticoagulant drug level is not enough, the level of stroke or embolism can be high. Conversely, when the level of drug is high internal bleeding may occur and coagulation process may not start after a serious injury. According to this condition, monitoring of blood coagulation for long term is part of an important haemostasis therapy.

Long coagulation time measurment usually indicate lack of specific factors in the blood. Various tests such as prothrombin time (PT), activated partial thromboplastin time (aPTT) have been developed to identify deficiencies of these specific factors [7]. Because thereby this procedure status of platelet aggregability will be determined. Irregularities in blood coagulation as noted previously plays important key role to bring the several diseases in the vascular system. Therefore, blood coagulation time measurement provides vital information regarding the patients.

In general, blood coagulation detection methods can be classified into four major groups:

### 3.2 Optical sensing

In this approach optical intensity is measured continuously. Optical properties (usually fluorescence) of the added reagents into the blood plasma change due to the chemical changes that occur during coagulation.

Typically, in this method light source and detector located on the different side of device. Figure 3 shows all parts of this device. The generated light by sours of light transmitted through blood samples and reflected, scattered or absorbed light from fluid can be measured by the detector. When clotting occurs the amount of the blood absorption changed. Hence, the value of detected light changes. This method is able to measure blood coagulation time for long term.

Spectrophotometer measures coagulation time by detecting optical intensity variations at different wavelengths. This approach is widely used in hospital or lab based instruments, but it requires a centrifuging step to separate plasma from blood. Figure 3 shows schematic drawing of a spectrophotometer, the device consists of a light source, a monochromator for providing wavelength selection device such as a prism, a location for loading the samples, a photo detector and data recorder [8].

In additional, another method that reported based on optic sensing, is a laser as a light source for producing the incident light and sample defined as a transparent. When the light passed from blood samples, it will record by utilizing a camera.

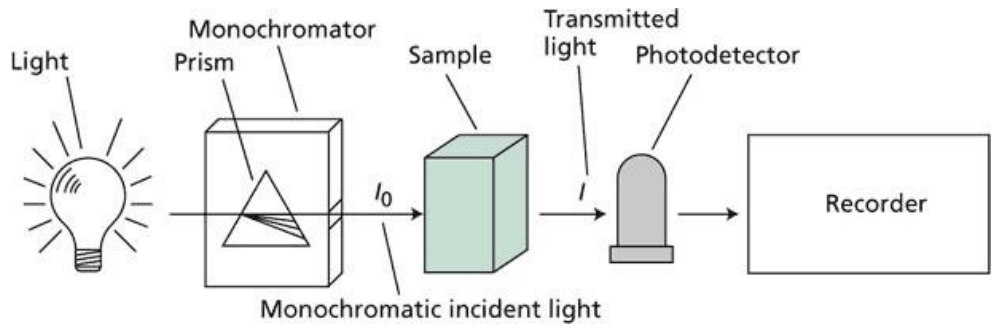

**Figure 3:** Simple schematic drawing of a spectrophotometer parts. Figure adapted from [9].

Eventually, speckle images which are recorded from top of the samples analysed for computing the intensity. During the clotting time image intensity changed. These changes show blood coagulation more details about this method for whole blood coagulation monitoring can be found in [10].

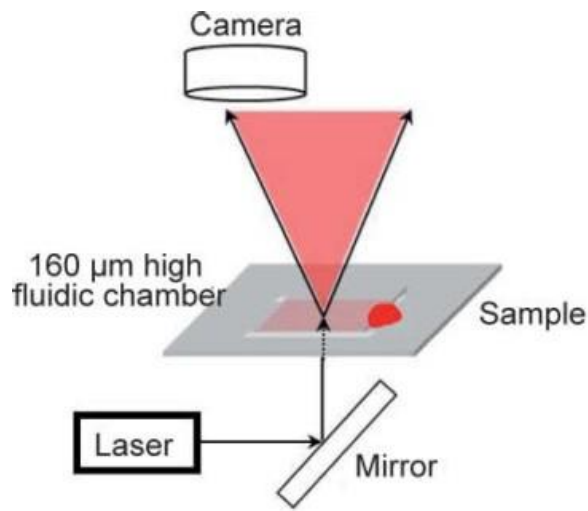

**Figure 4:** Schematic drawing of blood coagulation monitor for recording the images. Figure reprinted from [10].

## 3.3 Electrical sensing

These methods are based on detecting magnitude and phase impedance change of blood during clotting time. The most amount of resistance of whole blood comes from plasma resistance, cell interior and membrane of cells. Measuring the whole blood resistance can be provided invaluable information. Also by using this method sedimentation rates of erythrocytes and hematocrit can be measured. In current method there are two blood containers one of them is for reference measurement and other one to measure whole blood reaction with reagent during the time. This device includes two electrodes for each blood chamber. Further informations about fabrication and experimental results can be read in [11].

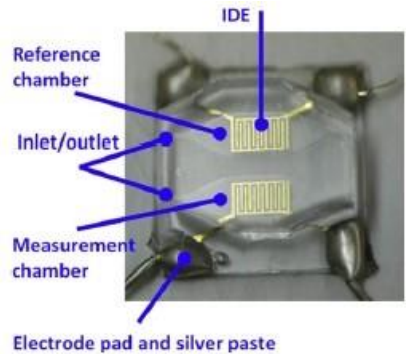

**Figure 5:** Top view of micro-fluidic channel for blood coagulation time measurement based on electrical impedance. Inter digital electrode (IDE) in this structure for electrical impedance sensing.
Figure reprinted from [11].

Another approach based on blood electrical properties has been introduced is used magneto-elastic sensors. The blood physical properties change during the clotting time. As noted in previous the final stage of coagulation process is changing the viscosity. The magneto-elastic sensors operate at a certain frequency by changing the viscosity, resonant frequency of sensors will shift. This happens accordingly equation

1.

$$\Delta f = -\frac{\sqrt{\pi f_0}}{2\pi \rho_s d}(\eta \rho_l)^{\frac{1}{2}} \quad (1)$$

where $\rho_s$ and $\rho_l$ are magneto-elastic sensor and liquid densities respectively, $\eta$ is viscosity of liquid, $d$ is magneto-elastic sensor thickness, and $f_0$ is resonant frequency of magneto-elastic sensor [12].

Since these procedures do not require large optical components and sample preparation, these methods are used in portable devices. However, both optical and impedance measurement approaches are indirect since they measure a secondary effect due to blood coagulation. Direct methods on the other hand measure the viscosity change of the blood during coagulation. So they are more dependable and immune to other factors such as high fat concentration that may affect optical and electrical properties of blood.

### 3.4 Mechanical sensing

Mechanical sensing methods are direct and they measure dynamics of a moving structure. Basically, micro or nano cantilever vibrates in the samples and there are two different methods for analyte binding by using the cantilever.

1. Static deflection Where bonding occurs on one side of cantilever, the imbalance appears in the strictures and deflection by this method is measurable.

2. Resonant mode Another way that is widely used for mechanical sensing is resonant mode, where binding occurs on cantilever surface proportional to particle mass the resonant frequency will shift.

Various methods can be appropriated for actuating or sensing cantilever motion, consist of mechanical, optical, electrostatic and electromagnetic methods. Cantilever based devices can be manufactured with different shapes and aspects using conventional MEMS technology such as bulk or surface micro-machining. One advantage of mechanical sensing based on cantilever is design the system for incorporation in microfluidic channel and portability for LOCs (Lab on a chips). By selecting the correct chemical reaction these systems are able to detect different chemical and biologic objects. Therefore, it is potential to manufacture as a large array for multi analytes. Figure 6 illustrates schematic drawing of a multi arrays based on cantilever for blood coagulation measurement [13, 14].

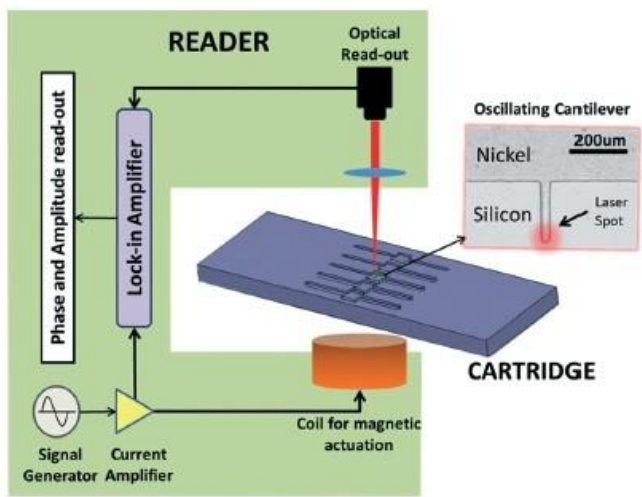

**Figure 6:** Schematic drawing of a cartridge based on micro-cantilever for multi analytes.
Figure reprinted from [14].

Another method has been introduced for blood coagulation monitoring is using a paper that whole blood transported on the paper for a constant time. In other words, when blood coagulated the viscosity of it changed and amount of the distance that it can travel on the paper reduced. The main advantages of this method are low cost because of using paper and this is simple method that complex set up is not required. The proposed method ("Blood coagulation screening using a paper-based microfluidic lateral flow device") and experimental data for different concentration of $CaCl_2$ depicted in figure 7 and figure 8 .

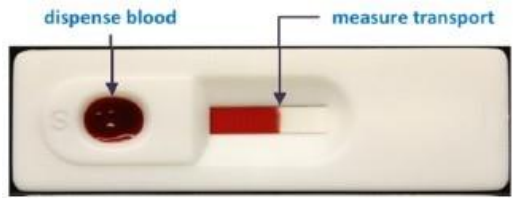

**Figure 7:** A photograph of an as-used assembled device (top) and of the device with the top piece removed to show the flow and device components.
Figure reprinted from [15].

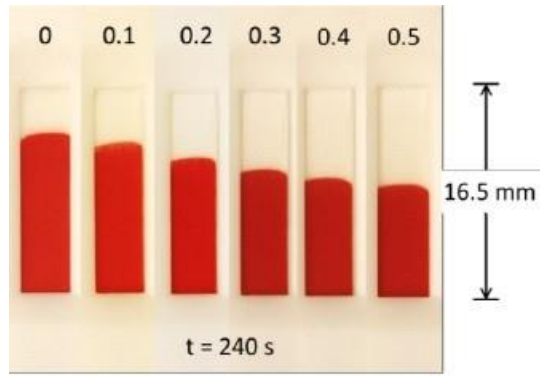

**Figure 8:** Photographs of LFA devices showing the transport of citrated whole blood during 240 s for several concentrations of added CaCl2 solutions.
Experimental data from [15].

## *3.5 Ultrasound sensing*

Methods using ultrasonic waves are also direct and they measure amplitude and phase change of ultrasonic waves propagating in a blood sample. Reflected acoustic waves from a sample are able providing invaluable information such as sound velocity, temperature and viscosity of blood.

In a typical arrangement an ultrasonic transducer is coupled to a container where blood sample is contained. The transducer is excited by short electrical bursts. The transducer creates acoustic waves in the buffer medium and in the sample. Acoustic reflections after reflecting from the various boundaries of the set-up reach back to the transducer where they are converted into electrical signals [16, 17]. Some important ultrasonic waves properties classified in below:

### 3.5.1 Acoustic wave propagation

A second order differential equation can describe the acoustic waves propagation in liquid medium which is called wave equation [18].

$$\frac{\partial^2 U}{\partial z^2} = \frac{1}{c^2}\frac{\partial^2 u}{\partial t^2} \qquad (2)$$

where $c$ is sound speed in liquid medium, $U$ shows generated displacement in z direction and $t$ denoted time. Also, similar equations is valid for pressure or density in liquid medium which are acoustic waves parameters. Solution of this equation brings $f(t \pm \frac{z}{c})$ that plus sign shows acoustic waves travel toward $-z$ direction and minus sign indicates acoustic waves propagation is in $z$ direction. This type of wave is called a longitudinal wave. The solution for equation 2 is

$$U = U_0 e^{j(wt-kz)} \qquad (3)$$

Where $w = 2\pi f$ angular frequency an $k$ is wave number and it's equal to $k = \frac{w}{c}$. For the longitudinal wave sound velocity is

$$c = \frac{1}{\sqrt{\rho G}} = (\frac{B}{\rho})^{1/2} \qquad (4)$$

Where $B$ is bulk modulus for liquid medium, $\rho$ is density and $G$ compressibility. $G$ defined negative change the volume by changing the pressure. The sound velocity in blood is similar with soft tissue and is $1540 m/s$.

### 3.5.2 Acoustic Impedance

Another important ultrasound characteristic is acoustic impedance in medium that's equal to

$$Z = \frac{P}{U} \qquad (5)$$

In this ratio $P$ is pressure and $U$ particle velocity in liquid medium. This equation is like Ohm law (ratio voltage to current). Therefore, acoustic impedance carries the same concept as the electrical impedance. The acoustic impedance can be rewritten as

$$Z = \rho c \tag{6}$$

The unit of acoustic impedance is $kg/m^2 sec$ or rayl. The acoustic velocity is dependence on temperature and it changes proportion to frequency.

### 3.5.3 Ultrasonic backscattering

The ultrasonic backscattering occurs where ultrasonic waves are incident on a diffuser with a smaller aspects than ultrasonic wavelength. This is a complex phenomena when scatter is a biological tissue with high attenuation. In other hand, this phenomena widely uses in pulse echo measurement specially in ultrasonic imaging.

Scattering cross section $\sigma_s$ can describe ultrasonic backscattering that comes from total power scattered by an object $P$ per unit incident intensity $I$, is given by

$$\sigma_s = \frac{P}{I} \tag{7}$$

Scattering cross section has a unit of $cm^2$ or $m^2$. Also, it can be expressed by flowing equation

$$\sigma_s = \int_{4\pi} \sigma_d d\Omega \tag{8}$$

Where $\sigma_d$ is differential scattering cross section and $d\Omega$ is differential solid angle.

Scattering cross section can be provided invaluable information about scattering strength of a substance. For an object with arbitrary shape compute the scattering cross section accurately is a critical computation. There are many approximation that could overcome to this problem. One of these approximations is Born approximation. In this method assumed incident waves and wave inside the abject are equal to each other. Generally, the Born approximation is useful when scatterer dimension is much smaller than ultrasonic wavelength or scatterers acoustic properties are similar to medium that scatterer located in that. More information about the Born approximation can be found in [19]. According to Rayleigh scattering theory, ultrasonic scattering strength is proportional to the 4th frequency power and 6th power of scatterer size. With considering the information noted in this section, scattering coefficient or volumetric scattering cross-section defines as an important ultrasonic parameter which is the power scattered per unit incident waves per unit scatters volume.

### 3.5.4 Ultrasonic attenuation

By passing the waves through an inhomogeneous medium, the waves power is reduced as function of distance due to the attenuation. The waves energy may be reflected, scattered or absorbed by the medium that waves incident it and result of this action appears as heat. Generally, the reflection and scattering by ultrasonic waves are referring same phenomena. Redistribution of energy when ultrasonic wavelength and wave front are smaller than object aspects called 'Reflection'. Respectively, when ultrasonic wavelength and wave front are greater than object size called 'Scattering'.

The pressure a monochromatic wave propagating in $z$ direction is reduced exponentially as function of $z$

$$p(z) = p(0)e^{-\beta z} \qquad (9)$$

Where $p(0)$ is pressure at $z = 0$ and $\beta$ is attenuation coefficient which is equal to

$$\beta = \frac{1}{z} \ln \frac{p(0)}{p(z)} \qquad (10)$$

The unit of attenuation coefficient is $nper/cm$ and also can be expressed in $dB/cm$. Due to frequency dependence of attenuation coefficient, when frequency is high the amount of the attenuation have to be high. Respectively, when the ultrasound transducers operate at low centre frequency the attenuation is low. In table 1 the important acoustic parameters summarized [20].

**Table 1:** Acoustic properties of different tissues.
Data collected from [20].

| Material | Velocity $m/s$ | Density $kg/m^3$ | Attenuation $dB/cmMHz$ | Acoustic Impedance $M\,Rayl$ |
|---|---|---|---|---|
| Air | 330 | 1.2 | - | 0.0004 |
| Blood | 1584 | 1060 | 0.2 | 1.68 |
| Bone, Cortical | 3476 | 1975 | 6.9 | 7.38 |
| Bone, Trabecular | 1886 | 1055 | 9.94 | 1.45 |
| Brain | 1560 | 1040 | 0.6 | 1.62 |
| Breast | 1510 | 1020 | 0.75 | 1.54 |
| Cardiac | 1576 | 1060 | 0.52 | 1.67 |
| Connective Tissue | 1613 | 1120 | 1.57 | 1.81 |
| Cornea | 1586 | 1076 | - | 1.71 |
| Dentin | 3800 | 2900 | 80 | 8 |
| Enamel | 5700 | 2100 | 120 | 16.5 |
| Fat | 1478 | 950 | 0.48 | 1.40 |
| Liver | 1595 | 1060 | 0.5 | 1.69 |
| Marrow | 1435 | - | 0.5 | - |
| Muscle | 1547 | 1050 | 1.09 | 1.62 |
| Tendon | 1670 | 1100 | 4.7 | 1.84 |
| Soft tissue (Average) | 1561 | 1043 | 0.54 | 1.63 |
| Water | 1480 | 1000 | 0.0022 | 1.48 |

*3.5.4.1  Blood coagulation measurement using the piezoelectric sensors*

The approach has been introduced to measure physical characteristics of sample fluids with using piezoelectric sensors. In this method, samples are loaded into a sample chamber in contact with a mechanically vibrating operating element. The vibrations are received by a piezoelectric sensor transducer and it is correlated to physical characteristic of a sample, such as viscosity or density. This method consist of the (annular) piezoelectric sensors and a circuit as a detector. Changes in electrical parameters such as a waveform, electrical resistance, voltage, resonance frequency or current of the transducer, due to viscosity or density changes during the clotting time detectable by the detector. The sensor has an actuator vibrating member that also functions as sensors element. Described devices in this section is able to monitor the blood coagulation for long time. Further details about the proposed method can be found in [21].

### 3.5.4.2 Blood coagulation measurement using the FBARs

Film bulk acoustic resonator (FBAR) is one type of acoustic resonators. Structure of this resonator includes a piezoelectric thin film sandwiched between two metal (most of the time gold used as electrodes) electrodes, the mechanical vibration excited at a specific frequency. The resonator generates an acoustic standing wave due to the acoustic isolation from the surrounding medium. Because of the small thickness of film, resonant frequencies range is approximately between 100 MHz up to 10 GHz. According to high quality factor (Q) in air, FBAR initially finds usages in the wireless communication circuits, such as radio frequency (RF) filters. FBAR can also be considered as a miniaturized version of QCM but can be exited in a different mode.

The basic difference between QCM and FBAR sensors is the resonant frequency quantity. QCM has a relatively low resonant frequency due to its geometry (thickness of it), thus a relatively low mass sensitivity. Mass sensitivity of these type resonators given by

$$S_m = \lim_{\partial m \to 0} \left(\frac{\partial f}{f_0}\right)\left(\frac{1}{\partial m}\right) \qquad (11)$$

Where $\partial m$, is the mass added to the mass sensor per unit area and $f_0$ is resonant frequency.

Other differences that can be considered are temperature stability and the Q factor. The FBAR devices quality factor changed to half of the original value where they operate in liquid environment. The same reduction occurs in QCM sensors where only 10% remains of the Q-factor. Also, the FBAR is preferred in temperature stability [22]. Fig.9 illustrates the schematic of the MEMS C-FBAR which is used for blood coagulation monitoring. The C-FBAR consists of a ring-shaped piezoelectric aluminum nitride (AlN) thin film that is sandwiched between the top and bottom gold electrodes; the ring is suspended and connected to the silicon substrate with one or more anchors. A drop of blood samples putted on top of this sensor and resonant frequency recorded continuously. More information about this method can be found in [23].

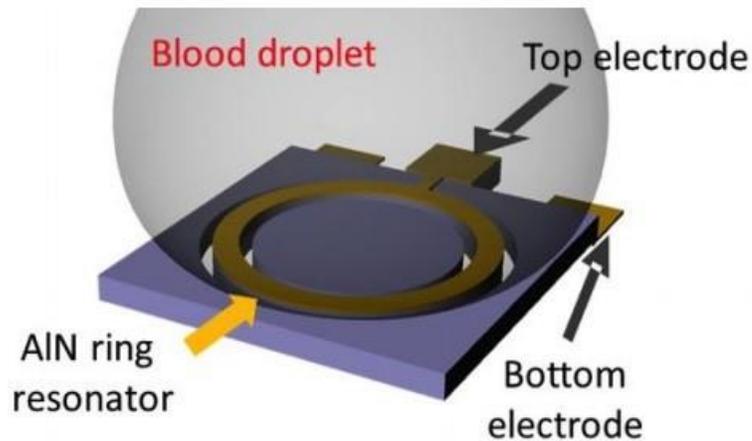

**Figure 9:** Schematic of a MEMS C-FBAR (contour-mode film bulk acoustic resonator) sensor with a blood droplet dispensed directly on top.
Figure reprinted from [23].

As mentioned previously, the purpose of this thesis is providing a portable system for measuring the blood properties such as blood coagulation in during the time. These devices for this aim have been introduced in recent decades, however, all devices need to huge blood sample and according to our knowledge they can not be categorized in portable devices due to required samples volume for processing and large set up for recording the data [24].

In the previous studies transducers with 5 MHz [16] and 30 MHz [49] have been used. The coagulation time has been measured by monitoring the amplitude of the reflections. The relatively low operation frequency of the transducer and the fact that the transducer is not integrated with the container necessitated using large volumes of blood samples between 4 mL to 40 mL. However, this prevented previously developed ultrasonic methods to be used in portable cartridge based systems. In this thesis we wll introduce a portable system that can perform coagulation time measurements using only 1 $\mu$L of blood which can be easily produced by patients from a finger prick by a lancet. To monitor coagulation time, high frequency (400 MHz) ultrasound is generated by transducers integrated with micro-channels to produce low cost cartridges. The proposed transducer in this thesis has maximum performance at 400 MHz. High operation frequency for used ultrasonic transducers could overcome on samples volume issue. In brief, main advantages of this device are low cost cartridges and low sample volume requirement thus, make the approach suitable for patient self-testing.

In the fallowing section we will present the cartridge geometry, used materials for reducing the cost and details of the piezoelectric transducer that uses a Zinc Oxide film. Also, we will explain the operation of the device by presenting pulse echo measurements. Finally, coagulation and anticoagulation reagent effect on whole blood with utilising the described sensor will display.

# CHAPTER IV

# PORTABLE BLOOD COAGULATION MEASUREMENT BASED ON ULTRASOUND

In this thesis very cost effective glass cartridges with integrated ultrasonic transducers were built. The cartridge provide a platform where one can perform various biological tests. One of advantages of this system is for coagulation tests 1 micro-litre of whole blood is enough to perform the measurement. Therefore, it is useful for patient self-testing. Due to system size it can be portable system. Materials and method for building the proposed sensor in this section will explain.

## *4.1 Cartridge*

The geometry of the cartridge is shown in figure 10. The cartridge is composed of two glass plates with various aspects. The top plate which is plate number 1 in figure 10 is a rectangular piece of glass that has 1 cm by 1 cm in width and length. The thickness of the plate measures 1 mm. The bottom plate (plate 2 in Fig.10) has larger lateral dimensions. The accurate size for plate is 2.5 cm by 1.7 cm. In our proposed system the transducers are located on top of the plate 2. As depicted in Fig.10 the thickness of plate 2 is 0.9 mm. The two plates are attached together by bio compatible double sided adhesively tape. The name of this special tape is 'Adhesive Research, Inc'. The thickness of the used tape is 75 micro meter. An L shape channel was manually cut using a razor blade into the top of the tape after it is placed on the bottom plate. The shape of the channel is arbitrary and in our set-up it was chosen to collect blood on the side of the cartridge to prevent it going over the electrical connections. The width of the appeared micro fluidic channel on top of the tape is approximately 1 mm. The cut is aligned with the transducers which are on the other side of the bottom plate. The top plate is placed on top of the tape such that two openings are created at both ends of the cut. The exposed holes provide inlet and outlet for the blood sample. Top view of plate 2 when the transducers are located on that depicted by Fig.11. In this figure, 5 transducers, sticky tape and microfluidic channel for loading the blood samples are obvious. Thus, very cost effective glass cartridges with integrated ultrasonic transducers were built.

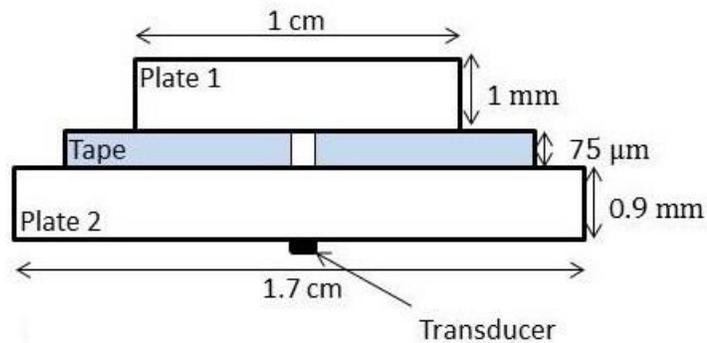

**Figure 10:** Side view of schematic drawing of a typical cartridge. The used materials in this cartridge are classes for providing the cost effective cartridge for self-testing purposes.

## *4.2 The ultrasonic transducers*

Figure 12 shows fabricated Zinc Oxide (ZnO) transducers. The fabrication process starts with patterning of positive electrodes on a glass substrate. The electrodes are made of 1500 Angstroms thick gold. 8 $\mu$m thick ZnO film is deposited over the electrodes using magnetron sputtering technique [25].

The last step is depositing and patterning of second gold layer for providing the ground electrode. At the end of the fabrication process, the transducers are formed at the intersections of positive and ground electrodes where Zinc Oxide film is sandwiched between the top and bottom electrodes. Further details of fabrication about these transducers can be found in reference [26]. The transducers, which are used in this research has 500 $\mu$m by 500 $\mu$m and the ZnO film thickness that sandwiched between two electrodes is 8 $\mu$m. The centre frequency of the transducers corresponding to this geometry is 400 MHz. In Fig.12 there are 5 transducers with mentioned aspects. In the experiments we just used the centre transducer for measurements by connecting the electronic connection to that. But, there are no limitation for utilizing of any of the 5 transducers.

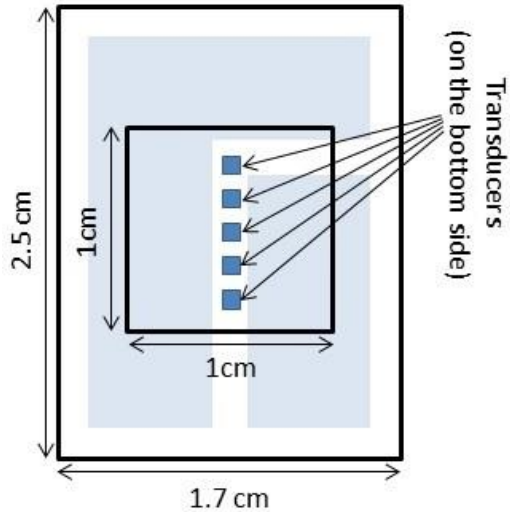

**Figure 11:** Top view of plate 2 in Fig.10 where transducers are located.

## *4.3 Electronics*

In this research for pulse echo measurements we built a custom RF system using off-the-shelf instruments and microwave components. Figure 17 shows a block diagram drawing of the all electronic parts for producing a high frequency signal. In the experimental set-up, RF signal generator (TTi-1GHz-TGR1040) that is able to generate 10 MHz to 1000 MHz frequency range generated 400 MHz sine wave. Power at the output setted at 0 dBm. However, this RF signal generator is able providing up to +7 dBm power at output. Figure 13 shows RF signal generator with electronic specification. The produced sine wave signal by RF signal generator was amplified using a microwave amplifier (Mini-Circuits-ZHL6+). Operating temperature for amplifier is -20°C to 65°C range and by applying 24 volte of the DC voltage it can be provided 21 ± 1.2 gain at out put. The amplified signal at out of the amplifier observed by oscilloscope and 20 dB gain for that detected.

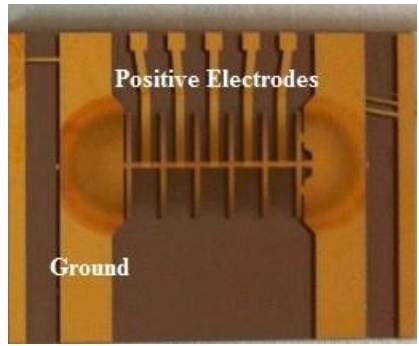

**Figure 12:** A photo of ultrasonic transducers. (Bottom side view of plate 2 shown in figure 10.)

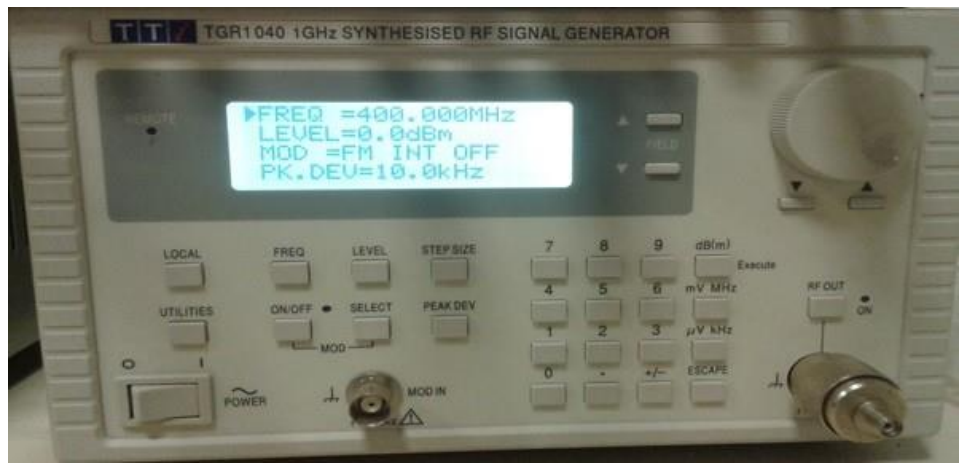

**Figure 13:** A photo of high frequency signal generator.

In the next stage, TTL switches (Mini-Circuits-ZYSWA-2-50-DR) were used to gate the continuous sine wave to generate a burst signal. Each switch needs to 5 volte DC voltage for operating and this type of switch has 1.5 dB insertion loss between DC to 500 MHz frequency rang. The electrical schematic illustrated in Fig14. .

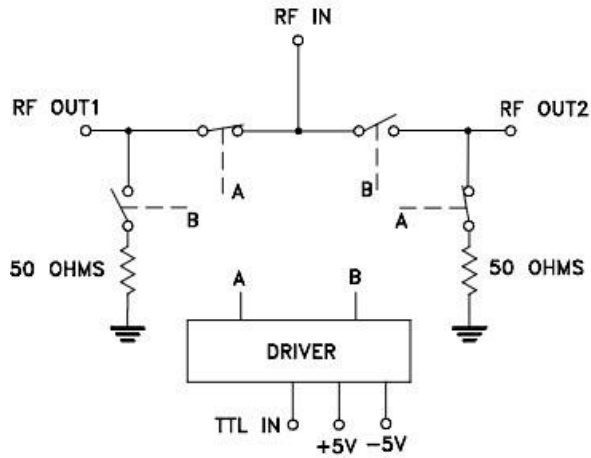

**Figure 14:** Electrical schematic of Mini-Circuits-ZYSWA-2-50-DR switch. Reprinted from Mini-circuit data sheet.

To increase the on/off isolation (approximately 60 dB isolation) two switches were used in series with same TTL signal. The control signal for the switches was generated by a function generator (Agilent 80 MHz function-33250A).

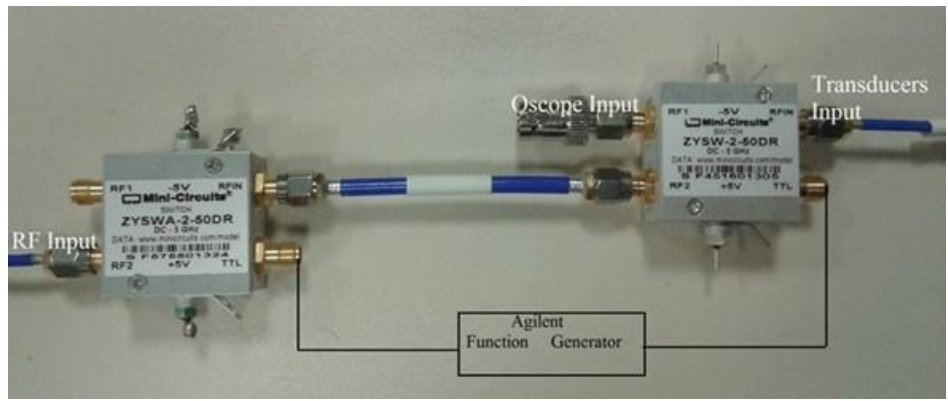

**Figure 15:** Two TTL switches connected in series for more isolation.

The burst length was variable from 10 nsec to 1 msec. In the our experiments, we used 50 nsec gate length. The high voltage sine bursts with very short widths (50 nsec) generated by this method, and they were applied to the transducer. The repetition rate was 1 KHz in the experiments Fig.16 depicted the Agilent function generator with electronic specification. In the transmit cycle the transducer is excited by the tone burst. In the receive cycle the transducer is connected to the oscilloscope (Tektronix-TDS654C) to detect the reflections coming from the transducer. The reflected acoustic pulses from the cartridge were sampled at 5 GHz using the scope. The digitized data were transferred to a personal computer using GP-IB interface for further processing.

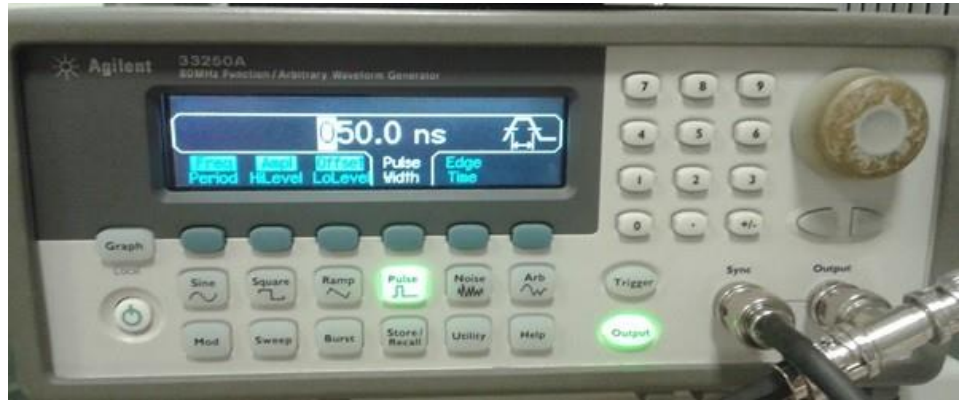

**Figure 16:** Agilent function generator for producing the control signal.

## 4.4 The temperature control system

The experiments were performed at a constant temperature. Typically, coagulation time measurements are done at 37°C degrees which is the human body temperature. Therefore, we designed a temperature control system to keep the cartridge temperature at 37°C ± 0.5°C. We placed the cartridge on a copper plate (7.5 x 4 x 1 cm) which has a temperature sensor (Dallas sensor DS18B20) at the centre and a heating wire (Fig.19). The current for the heating wire was provided by pulse width modulated MOSFET transistor (IRFZ44N). The modulation frequency was controlled by

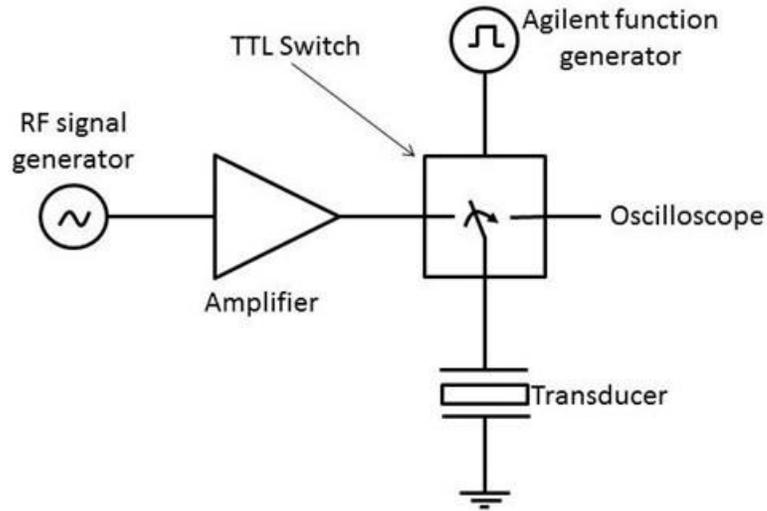

**Figure 17:** Block diagram of the set-up used for pulse echo measurements

Arduino board. A PID controller was also implemented on the Arduino board to control the temperature of the copper plate. Fig.13 shows the temperature control system parts.

### 4.4.1 PID parameters values calculation

As noted in previous paragraph for providing the constant temperature we used a PID which is uploaded on Arduino board. First of all, heat exchange transfer function in copper plate calculated experimentally. According to this purpose, by applying a DC voltage to the heat wire (12.5 volte), the temperature inside the copper plate sweeps from room temperature (25) up to 75 °C . This is a method for computing heat transfer time response in copper palate. In the next step of transfer function computing, constant time and dead time with considering the time response of copper plate calculated.

$$\tau = \frac{3}{2}(T_2 - T_1) \tag{12}$$

$$a = T_2 - \tau \tag{13}$$

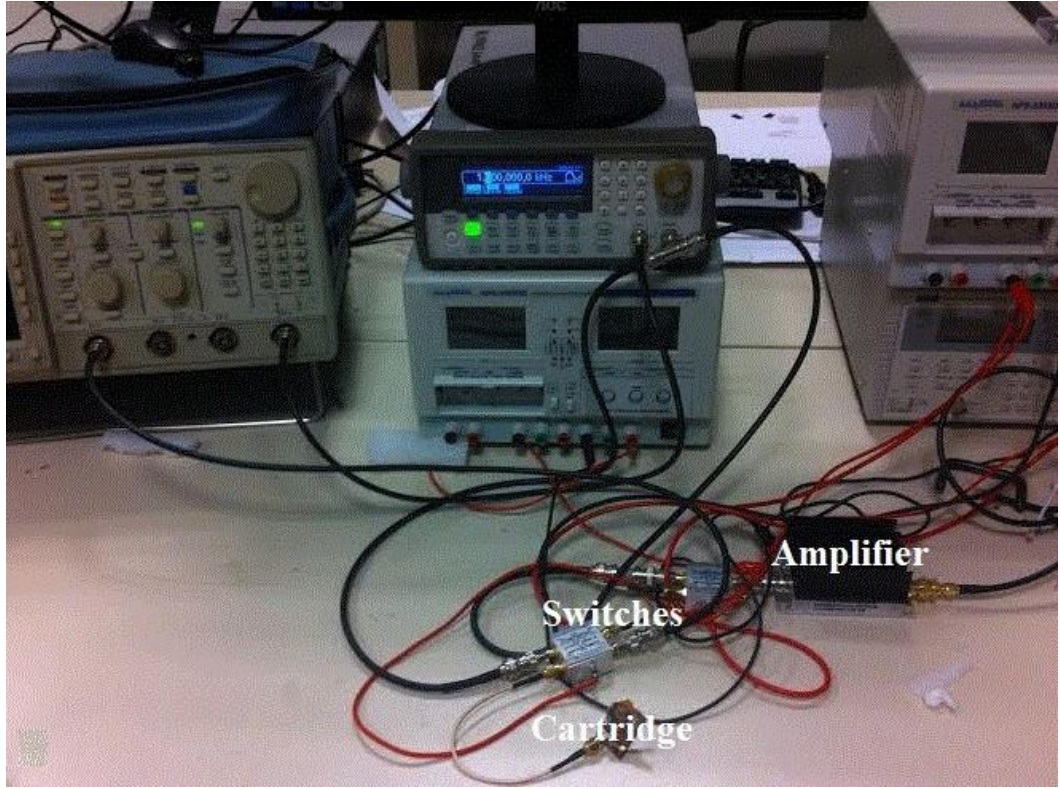

**Figure 18:** A photo of the measurement set-up.

Where The values of $T_1$ and $T_2$ are the times where the response attains 28.3 % and 63.2% of its final value, respectively. Also, $\tau$ is constant time and $a$ is dead time according to heat transfer time response. After calculating transfer function for copper plate (Equation 14), its imported to MATLAB for computing the PID parameters for 20% overshoot and without any steady state error.

$$G(s) = \frac{e^{-as}}{\tau s + 1} = \frac{e^{-14s}}{21s + 1} \tag{14}$$

In this system 20 % overshoot is acceptable for preheating the transducers and cartridge, then amount of the damping ratio $\xi$ is equal to 0.456, it given by

$$OS\% = e^{-\frac{\sqrt{\xi\pi}}{1-\xi^2}} * 100 \tag{15}$$

After several mathematics steps, with considering the root locus of temperature control system that illustrated in fig.22 , proportional gain $K_p$, derivative gain $K_D$ and integral gain $K_I$ for fallowing equation calculated. Moreover for more accuracy we used MATLAB PID automatic optimization and PID values double checked.

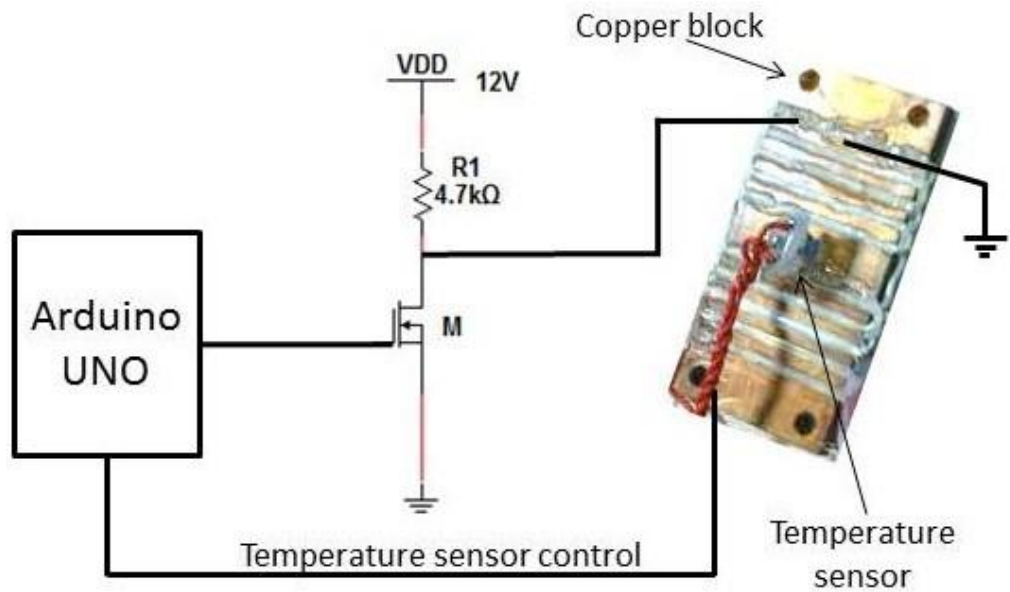

**Figure 19:** Schematic drawing of electronic circuit of the temperature control system for keeping the temperature at 37°C ± 0.5°C.

$$G_{PID}(s) = K_p + \frac{K_i}{s} + K_D s \tag{16}$$

By applying the described PID to the system both settling and rise time improved and we could reach to 20% overshoot. Step response of this temperature control system after applying the PID shown by figure23 and steady-state error depicted by Fig.24 .

### 4.5  Pulse-echo measurements

Fig.25 shows a simplified drawing of the acoustic reflections in the cartridge, in this figure where R1 is the first reflection from glass substrate and P1 is the first reflection from the channel top when the channel is filled with a fluid. The transducer is excited by a short tone burst. The short tone burst produced by custom RF system that explained in Electronic section. The electrical signal is converted into an acoustic pulse in plate 2 by the transducer. After propagating along the glass plate (Plate 2), the acoustic pulse reaches to the channel/glass interface.

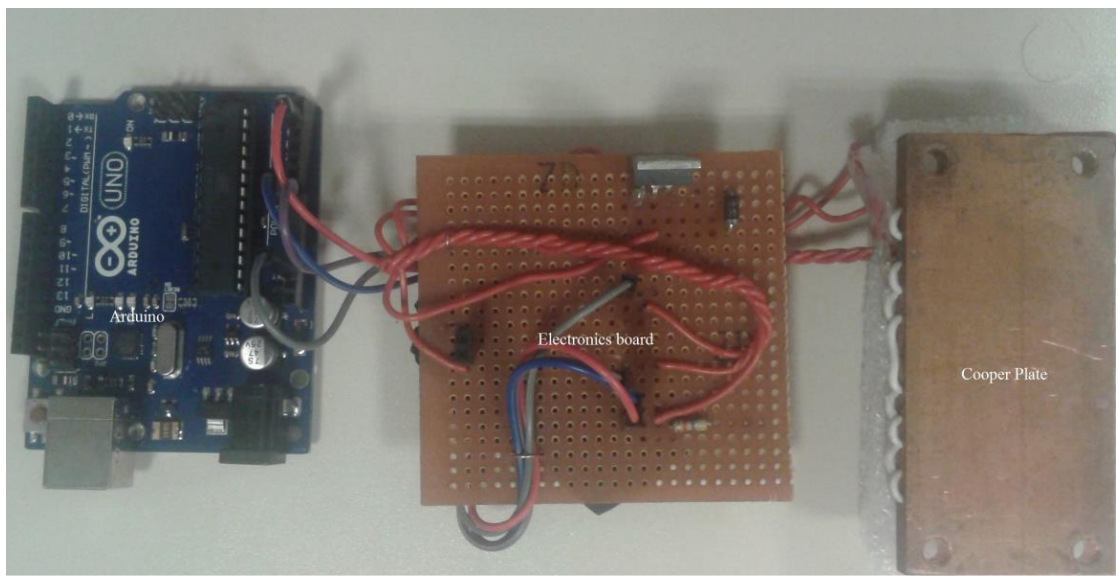

**Figure 20:** A photo of the temperature control system parts.

If the channel is empty most of the pulse is reflected back. The reflected pulse then arrives back to the transducer where it is converted into an electrical signal. At the same time, some portion of the pulse reflects back into the plate again. The back and forth reflections of the acoustic pulses are shown in figure 26. The first reflection is labelled R1 in Fig.26 and the consecutive reflections are labelled R2, R3 and so on. When the channel is not full with liquid sample, the acoustic waves are reflected in the glass substrate (Plate 2 in Fig.25) and they decay can be observed due to the acoustic loss and diffraction loss. The thickness of the plate is 0.9 mm and assuming the longitudinal sound velocity in glass is 5600 m/sec, the first pulse should arrive around 345 nsec which is in line with the time delay between the reflections shown in figure 26. Fig.27 shows the reflections when the microfluidic channel is filled with a fluid. Initially, the channel was filled with DI water in the experiments. In this case, some portion of the acoustic pulse transmits into the liquid at the first glass/channel interface. The acoustic pulse propagates through the liquid and after reflecting from the top of the channel and propagating through the liquid and the plate, it reaches to the transducer. The additional pulse (P1) can be observed on the scope screen as shown in Fig.27. The consecutive multiple reflections in the channel are labelled P2, P3. The channel height is 75 $mu$m and the velocity of sound in water is 1500 m/sec, therefore P1 reflection should arrive 100 nsec after R1, and this fact is obvious in Fig.27.

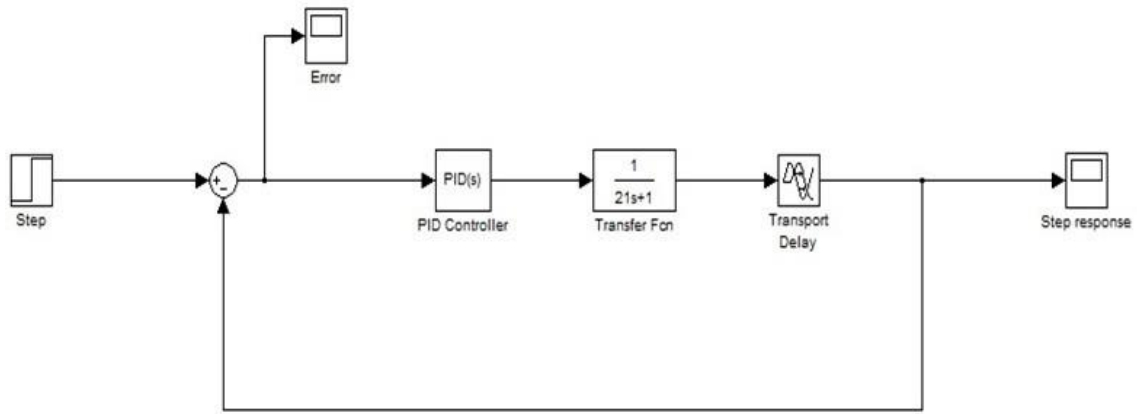

**Figure 21:** Set-up for simulating the temperature control system in MATLAB.

## *4.6 Data collection*

Data collection was achieved through LabVIEW and MATLAB, for this purpose we used GP-IB (National Instruments-USB-HS) for making a connection between oscilloscope and computer. The data collected from screen of oscilloscope with a simple graphical code that wrote in 'LabVIEW' software. The total data imported to MATLAB for computing the Root-Mean-Square of reflection amplitude. RMS values

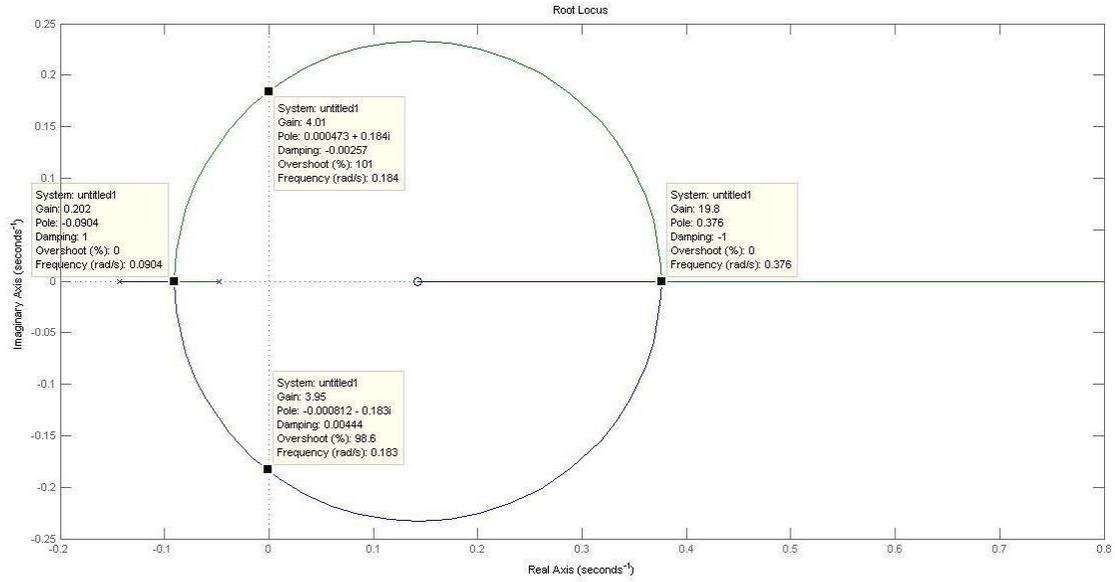

**Figure 22:** Root locus of temperature control system by using MATLAB.

computed with fallowing equation

$$V_{RMS} = \sqrt{\frac{1}{N} \sum_{n=1}^{?} |V_n|^2} \qquad (17)$$

Figure 28 shows the reflection (P1) amplitude over 600 seconds. For this measurement, oscilloscope traces were acquired at every 1 second and stored in a computer for 600 seconds. Then, RMS amplitude of P1 was measured over a 30 nsec window which is placed to cover P1 reflection between 600 nsec to 630 nsec. The slight drift on the curve is due to the temperature equalization of the water.

## *4.7 Blood sample preparation*

Human blood samples were drawn from a healthy volunteer and stored in a 250 mL clinical tube with 0.9 percentage citrate (BD Vacutainer 9NC 0.109M). For all the experiments, fresh blood samples of less than one hour old were used. Before putting the samples inside the channel, blood samples, reagent solutions and the channel were preheated for 2 minutes at 37 degree Celsius by the temperature control system. After preheat, 50 µL of blood samples were mixed with reagent solutions for specified seconds outside of the microfluidic channel. Finally, 1 µL mixture solution was applied to the inlet opening of the microfluidic channel for measurements. Also, required sample volume for doing the experiment can be drawn from fingers head of the hand by a lancet easily and painlessly because of very low blood sample volume required. This method can bring the fresh blood for self testing.

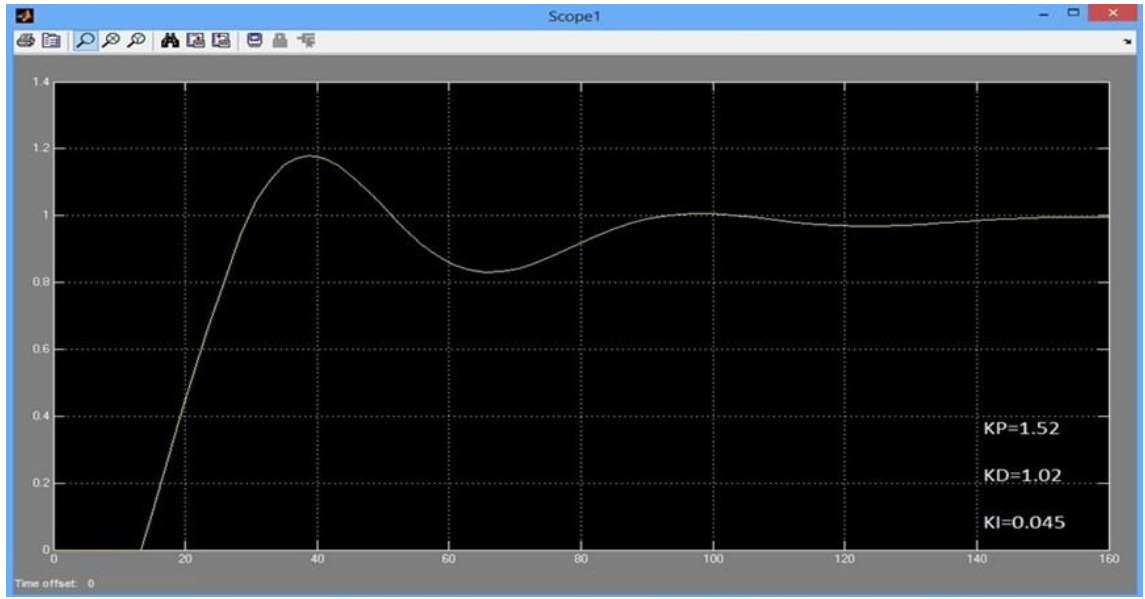

**Figure 23:** Step response of temperature control system with PID.

## *4.8 Experimental results*

Figure 29 shows a typical pulse-echo result when the channel is filled with blood. The multiple reflections in blood are labelled P1 and P2. P2 reflection is barely visible. Due to the high acoustic attenuation in blood [20] the amplitude of the multiple reflections decay much faster compared to the case when the channel is filled with water. Channels with small channel heights result in higher reflection amplitudes. But the height should be high enough to allow well separated pulses in the time domain such that one can easily measure the amplitude of the P1 pulse without interference from R1 reflection. To observe coagulation, the amplitude of the P1 pulse is measured by calculating the RMS value of the oscilloscope trace between 600 and 630 nsec.

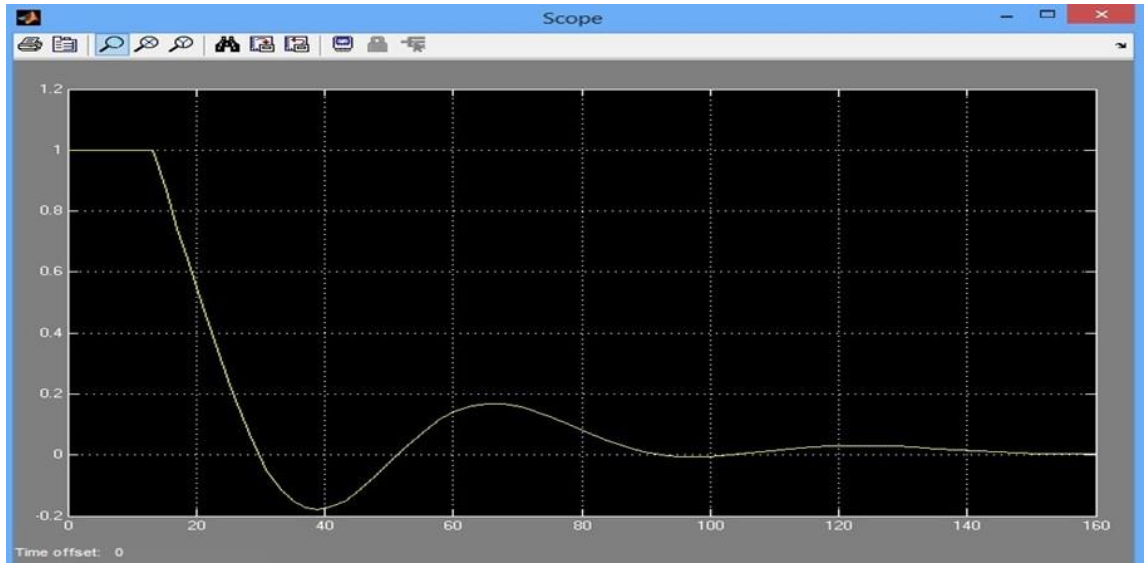

**Figure 24:** Steady-state error of the temperature control system. By applying the PID to the system after around 100 second, the system steady state error is zero.

The method was tested first with citrated blood. The channel was filled with 1 µL of citrated blood using an auto pipette following the procedure explained in the previous section. Fig.29 shows the amplitude of the P1 pulse measured over 900 seconds. We hypothesise that the fluctuations in the measurement are due to the slow sedimentation of red blood cells. For this experiment, we did not observe any sudden change of the amplitude. This indicates the blood did not coagulate in the channel. We also observed citrated blood outside the channel in a small container and as expected the blood did not coagulate over 900 seconds, either. In the second experiment, we mixed blood with 0.25M calcium solution $CaCl_2$ outside the channel for 20 seconds.

We applied 1 µL of the solution to the channel using an auto pipette. Measurement of P1 amplitude with calcium is shown in the same figure. There is a sudden change on the amplitude around 720 seconds which is due to the coagulation. Simultaneously, we also observed coagulation of the blood that we kept outside the channel. In the next experiment, we mixed blood with calcium solution and aPTT coagulation agent (SynthASil) for 20 seconds outside the channel. We loaded 1 μL of blood in the channel and observed the amplitude of P1 pulse. The result was also plotted in Fig.30. Around 75 seconds we observed a sudden amplitude change. Accordingly, the sample outside the channel also coagulated.

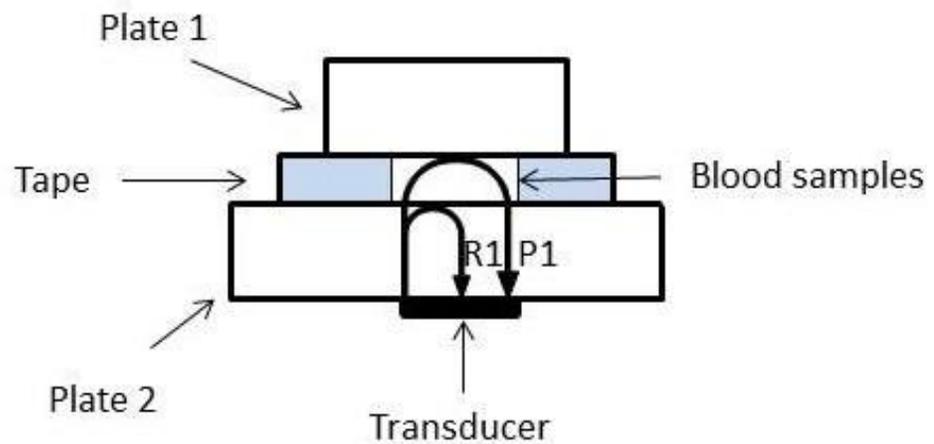

**Figure 25:** Simplified drawing of acoustic reflections in the cartridge.

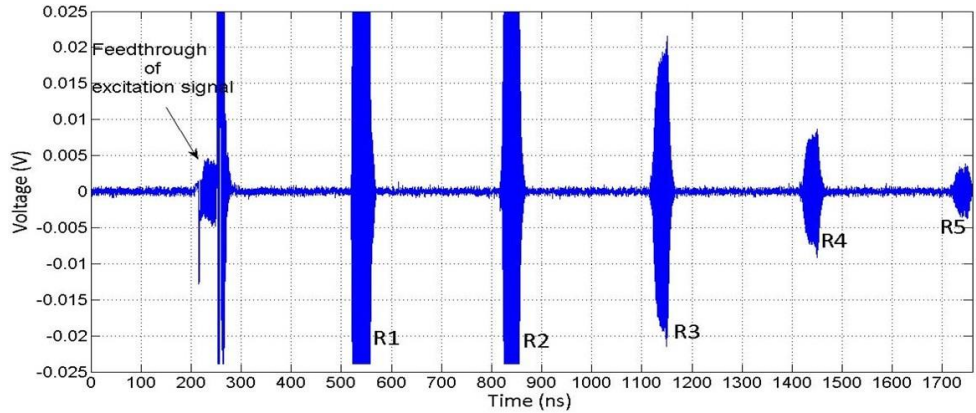

**Figure 26:** Reflections in the cartridge, when the channel is empty, the acoustic waves are reflected in the glass substrate (Plate 2 in Fig.5) and they decay due to the acoustic loss and diffraction loss.

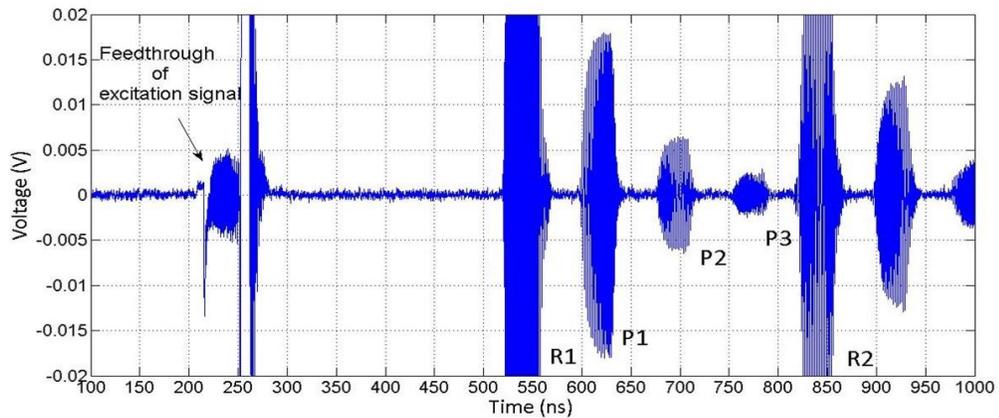

**Figure 27:** The microfluidic channel is filled with DI water and the measurement was done at 37°C.

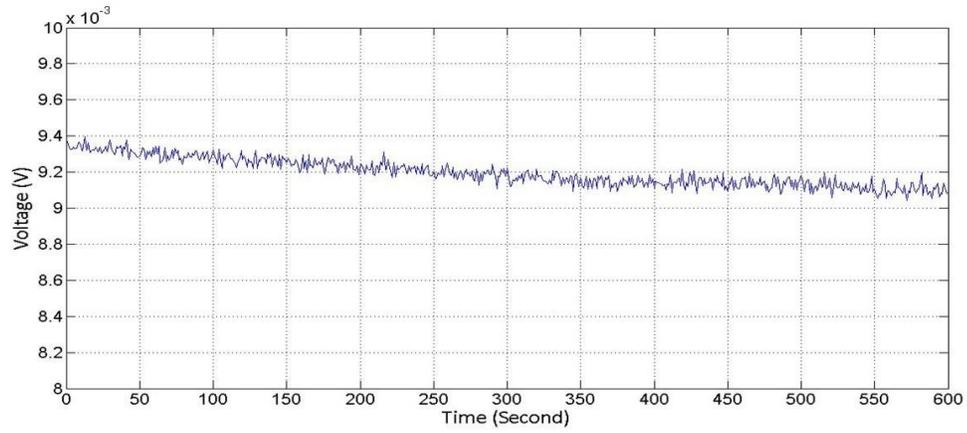

**Figure 28:** RMS value of P1 when the cartridge is filled with DI water.

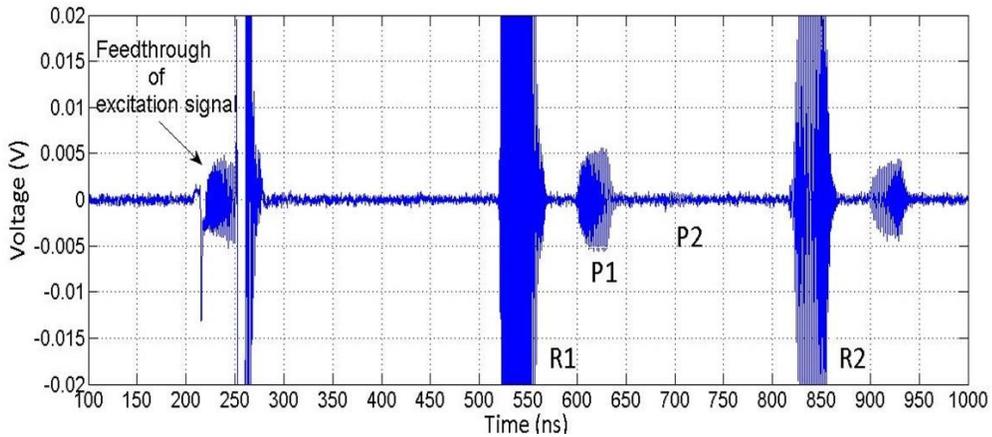

**Figure 29:** The cartridge is filled with 1 µL of whole blood sample without any reagents and measurement took place at 37C. P1 is the first reflection from the top of the channel. The multiple of this reflection, P2, is barely visible.

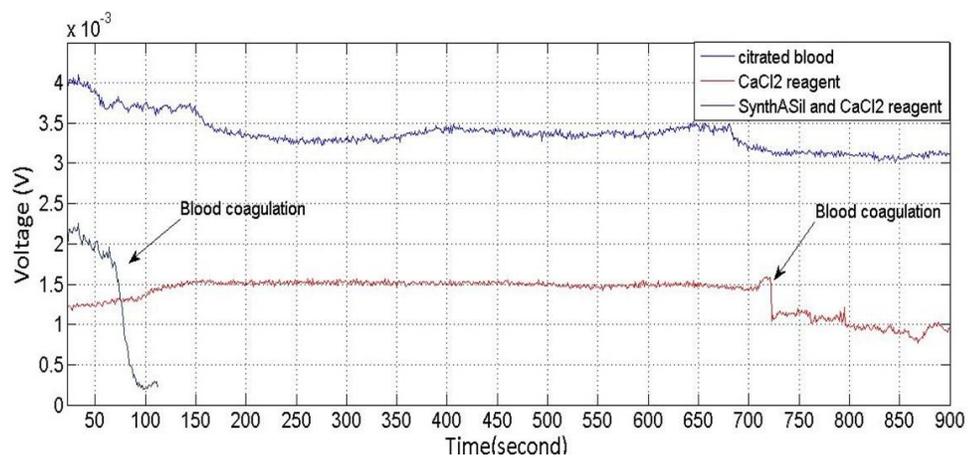

**Figure 30:** Coagulation data. The RMS amplitude of P1 between 600 and 630 nsec was recorded. Citrated blood data has been collected to check the stability of the method. Blood with CaCl2 solution exhibits coagulation around 720 seconds. Blood with aPTT reagent coagulates much faster (around 70 seconds) as expected. Note that there is also 20 sec. mixing time before the blood is applied to the channel.

# CHAPTER V

# RESONANT MASS SENSORS BASED ON CMUT

## *5.1  Introduction*

Resonant mass sensors are widely used for sensing applications due to their high sensitivity and low noise floor. The noise floor of the sensor is determined by the quality factor of the resonator. High quality factor results in a reduction of the measurement noise. For immersion applications in liquid, the quality factor of the resonating structure is reduced by the fluid loading which results in increased noise floor. The objective of this part of thesis is to offer a solution for low quality factor problem of resonant sensors where they operate in liquids. The proposed solution uses membrane shape resonators which are can be fabricated by CMUTs technology.

## *5.2  Molecule detection in air*

Detection of minute chemicals has many scientific and industrial applications. In gaseous environment chemical detection instruments in general are called electronic noses. Electronic noses are successfully demonstrated to detect various fragrances for cosmetic industry [27], to separate spoiled food and beverages [28], for environmental monitoring [29] and to monitor explosives [30]. It is possible to expand the list of applications to include medical and process control fields. Due to wide application range, several dozen companies are now developing and selling electronic nose devices globally. Odotech, eNos and  etc.

There are several sensor technologies that are used in the sensor arrays. These include optical detection [31, 32, 33], thermal or calorimetry [34], electrochemical detection[35], surface acoustic wave [36] and resonant mass detectors. The latter technology attracts more interest among the others. Resonant mass detectors are relatively simple and they can achieve very high sensitivities. But the main reason they have gotten more attention is that they can also work in immersion. Therefore they are the method of choice for most biological applications.

## 5.3  *Resonant mass sensors*

Resonant mass sensors can detect nano-sized particles to individual molecules without any labelling. Labelling increases the cost and complexity of the tests therefore it is not desirable. These sensors contain mechanical structures that vibrate with constant amplitude. In biological applications, the structure is vibrated in aqueous solutions.

Resonant mass sensors detect small changes in the resonant frequency of the mechanical structure due to the additional mass of the target particle that binds to the sensor surface. When binding occurs, the resonant frequency drops proportional to the mass of the target analyte. One can convert the frequency drop to the additional mass using the following equation

$$\frac{\Delta f}{f} = -\frac{1}{2}\frac{\Delta m}{m} \qquad (18)$$

Where $f$ is the resonant frequency and $m$ is the modal mass of the vibrating structure. More accurate calculations require FEM (finite element method) analysis but equation 18 provides fairly accurate results if the vibrating structure can be modelled by a simple harmonic oscillator.

As for the mechanical resonator, various mechanical structures have been proposed for detecting chemicals. In the earlier days, QCM (Quartz Crystal Microbalance) became the natural way for detection [37]. QCM has long been used for measuring thin film thickness for deposition instruments. So particle detection was a natural extension of this application. Due to its energy trapping feature, QCM does not couple well to the fixture that is holding the crystal. Therefore, it is possible to achieve very high quality factors (100000 1000000) in air using this method. In immersion, the crystal oscillates in shear mode where the crystal surface move parallel to the liquid and hence QCM does not couple well to the liquid, either. Because of this, QCM still achieves very high Q in immersion. But although it is possible to achieve very high Q operation and hence low noise for frequency monitoring, the sensitivity of the QCM is low due to its relatively large mass. QCM based sensors are also not suitable for array operation as well as not suitable for applications where small amounts of specimen are available. In addition, QCM requires piezoelectric materials which cannot be machined using standard high volume production techniques such as CMOS. This makes them difficult to integrate with electronics and therefore prevents building large arrays. Also these sensors are difficult to operate in liquid environment due to requirements of passivation of electrical connections.

To solve problems associated with QCM, cantilever based sensors have been proposed and they became widely used [38, 39]. Initially, cantilever based sensors were used in air. Later their operation in liquid has also been demonstrated [40].

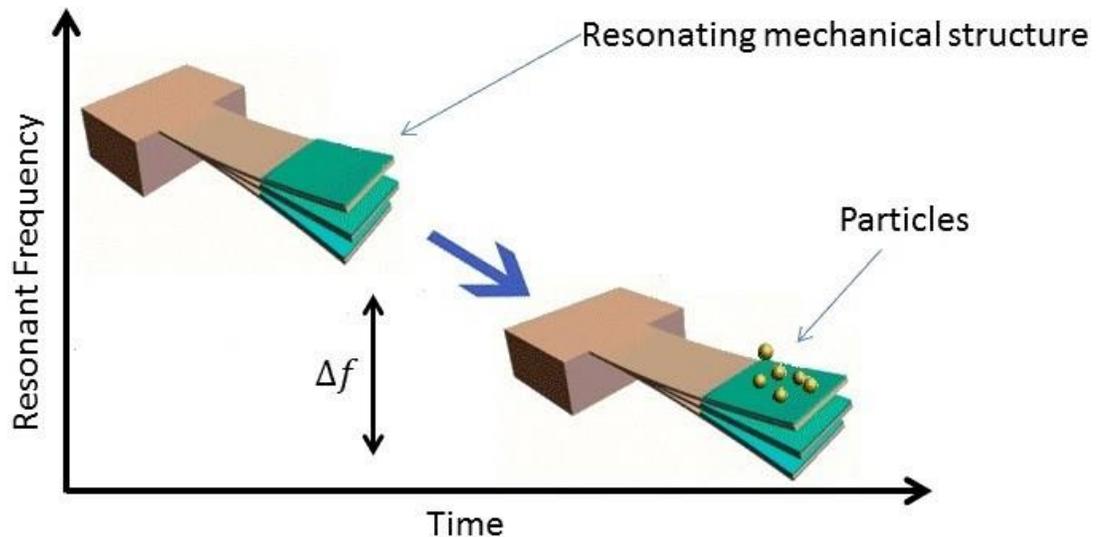

**Figure 31:** Resonant mass sensor structure based on cantilever. Figure adapted from [41].

Where the micro-cantilever operates inside the liquids, quality factor and resonant frequency both of them drop proportional to viscosity of liquids. Figure 32 illustrates an example for a cantilever that is operates in two different environments. Some approaches for overcoming the low quality factor of cantilever in liquids problem presented. One of them could increase the Q factor for a cantilever up to 600 by using an electronic feedback(further details about this method can be found in [43]), however, parallel with this increasing the noise floor increased.

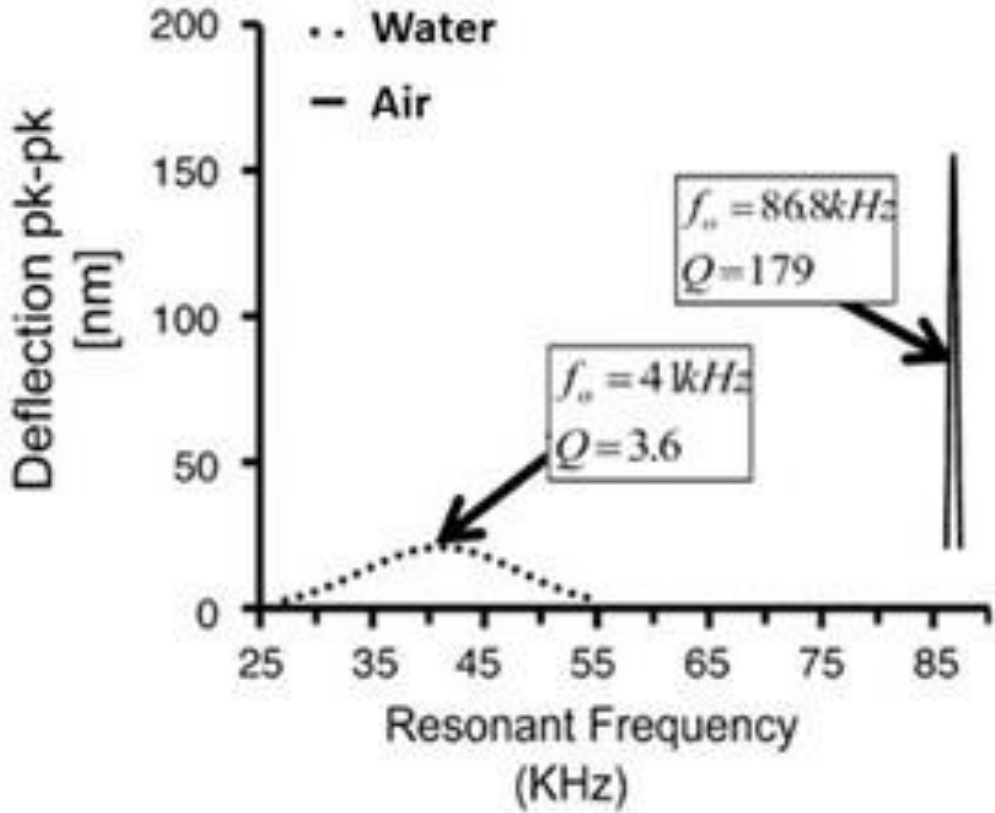

**Figure 32:** Resonant frequency and quality factor of a cantilever based on sensor where is vibrating in water and air.
Figure reprinted from [42].

In summary, there are couple of disadvantages of cantilever based sensors. For example, functionalization of large cantilever arrays is difficult. There are various methods have been proposed such as dipping the cantilevers in microfluidic channels where each channel carries a different functionalization fluid. These methods are fairly complex and are not suitable for volume production. Also, different stress levels of the cantilevers make functionalization problematic due to different deflection profile at each cantilever. Another drawback of the cantilever based sensors is that the optical beam should travel through the liquid that is being measured. In biological specimen optical path may get blocked by the large cells and other biological particles. Also, the refractive index changes of the biological fluids may also affect the measurement results. In addition, lack of integrated optical sensors limits the portability of the sensors based on cantilevers. Thus we can not use cantilever based on sensors as portable devices.

## 5.4 Resonant mass detectors based on CMUT

In this section of thesis, to solve the mentioned problems and improve the sensitivity, a CMUT (Capacitive Micromachined Ultrasonic Transducer) based sensor is proposed. In this sensor the mechanical structure is a membrane rather than a cantilever beam.

In the earlier work CMUTs were used as a resonant mass sensor in the air[44]. CMUTs were originally developed for medical and underwater ultrasonic imaging [45]. A CMUT typically consists of a large number of vacuum-backed resonating membranes connected in parallel. These transducers have inherent performance and manufacturing advantages that make them attractive devices for use in sensing applications compared to cantilevers. The major advantage is that it is possible to build higher resonant frequency structures with low mass. According to Equation 18 higher resonant frequency results in higher mass sensitivity. In addition, the membrane surface provides a flat plane where functionalization can be easily carried out. This enables building massive arrays that can detect different chemicals. Figure 33 shows optical image of the CMUT with several membranes cells, also in this figure top electrode demonstrated.

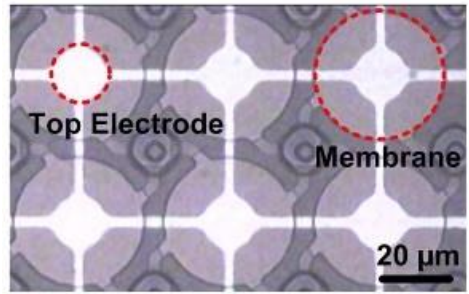

**Figure 33:** Optical image of the CMUT array with several membranes. Figure reprinted from [44].

If the sensor shown in Fig.33 is immersed in liquid, it will result in very low Q of around 1. Basically, the introduced method in this thesis is similar with traditional resonant mass sensor based on CMUT with some changes. Instead of connecting many membranes in parallel as shown in Fig.33, there will be only one membrane (active membrane) as the sensing element. There will be also other membranes around the active membrane. The non-active or passive membranes will create a pressure release boundary. The pressure release boundary will reduce the real part of the radiation impedance; therefore the coupling into the immersion medium will be reduced and quality factor of the active membrane will be increased. The geometry of the proposed device is shown in Fig.34 and Fig.35. Basically, both proposed geometries are same just the difference is in number and pattern of the passive membranes. The active membrane will be actuated by the top and bottom electrodes. If a sinusoidal signal is applied to top and bottom electrodes the membrane vibrates. Additionally, DC voltage as bias voltage is required. where active membrane able to vibrate and non-active membranes provide the pressure release condition to achieve high Q.

There is a correlation between acoustic losses and real part of the acoustic impedance. In other words, when acoustic impedance amount in the liquids medium is high, the value of Q factor will be low. In figure 36 real and imaginary parts of acoustic impedance in liquid with different boundary condition for a frequency range between zero to 40 MHz calculated analytically.

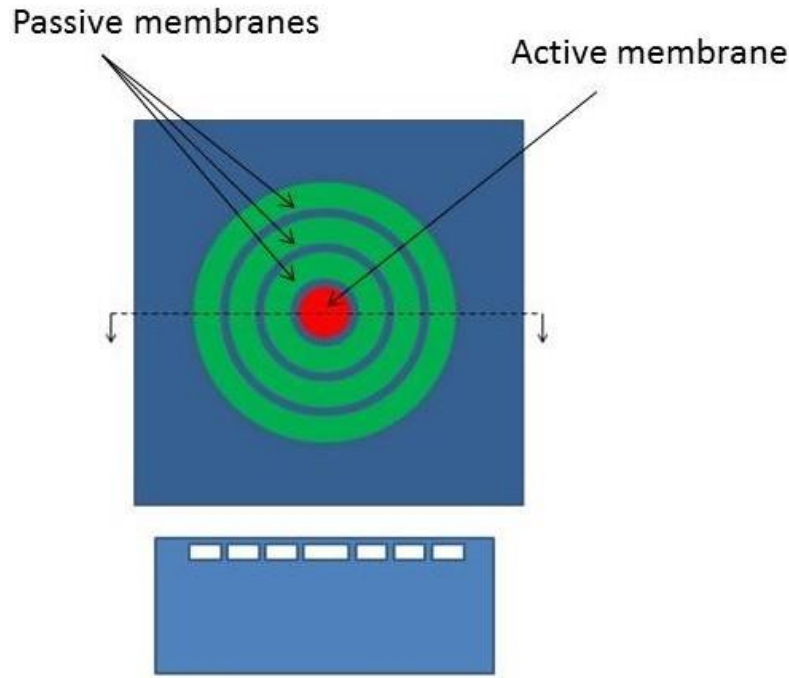

**Figure 34:** Resonant mass sensor with annular shape.

## 5.5 *Finite element model*

As a first step, a finite element model of the membrane will be built using a commercially available software package (ANSYS). Fig.37 shows one of the FEM models that was developed in the initial work. The membrane is 20 micro-meter in diameter and 1 micro-meter thick. The membrane material is single crystal silicon. We used "plane42" element for this membrane and silicon properties (Density=2332$Kg/m^3$ , Modulus of Elasticity=1.69 $*$ $10^{11} Pa$ and Poissons ratio=0.29) defined. Moreover, element "Fluid29" used for providing the liquid such as water. therefore, water properties (Density=1000$Kg/m^3$ and ultrasonic velocity=1500 $m/s$ ) inserted in software. The absorbing boundary in our simulation provided by type of element interface

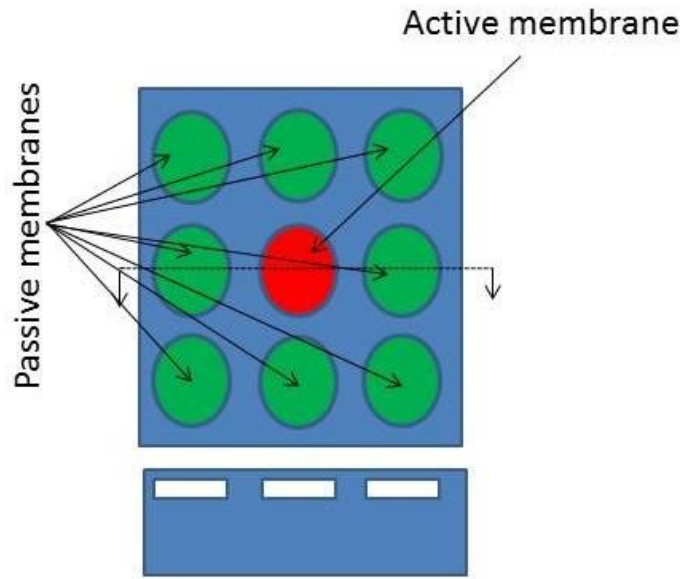

**Figure 35:** Resonant mass sensor with .

"Fluid129".

The quality factor of the oscillation is found by sweeping the harmonic frequency. Fig.38 shows the membrane displacement. In this study, two different boundary conditions were applied to the region next to the membrane. These are rigid baffle and pressure release boundary conditions. In rigid baffle boundary condition, the vertical displacement is assumed to be zero where as for pressure release boundary condition, the liquid particles are assumed to move in a pressure free field. The quality factor when the membrane operates in rigid baffle is 9.45. However, if the rigid baffle is replaced by a pressure release , Q increases to 440. Usually rigid baffles are easy to implement; just the solid surface will create a rigid baffle around the membrane.

One way of implementing a pressure release baffle is to build adjacent membranes to the original membrane as shown in Fig.39. Same analysis was repeated using the structure shown in Fig.39. The frequency sweep is shown in figure 40. In this case the Q is 350 excluding liquid viscous losses which will lower the Q 20%. There are also other resonances between 1 to 2 and 5 to 8 MHz due to the cross talk between the non-active membranes of the structure in the baffle. These initial results clearly show Q enhancement using passive membranes.

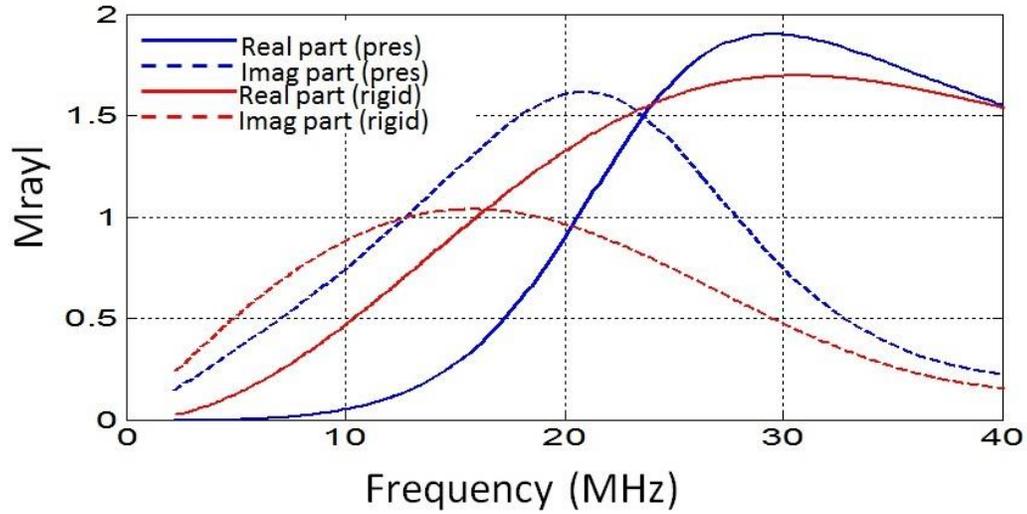

**Figure 36:** Real and imaginary parts of acoustic impedance values for pressure release and rigid baffle conditions.

Once again previous process performed on proposed sensor structure. In this step water viscosity considered. Therefore, we replaced previous material (Fluid29) to "Fluid79" element which is able to provide the viscosity for liquid. Then 1 cP viscosity added. Figure shows two process, one of them the viscosity of water ignored and other one viscosity of water considered. By adding the viscosity the quality factor around 20% drops. However, it is still more than the obtained quality factor in rigid baffle condition. The quality factor values given by

$$Q = \frac{A}{\sqrt{2}} \tag{19}$$

Where *A* is maximum average displacement that occurs at the resonant frequency.

Other effect of viscosity on proposed sensor that observed from simulation result is reducing the displacement. As shown in Fig.41 when viscosity of water added to the fluid properties, the average displacement around 15% drops.

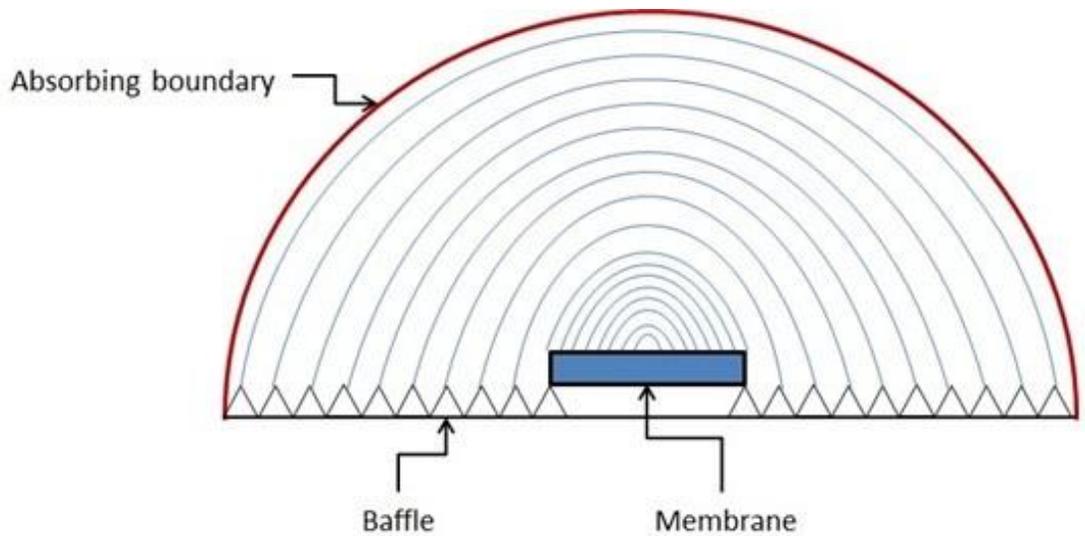

**Figure 37:** FEM model of membrane in ANSYS.

In summery, described sensor in this thesis is able to vibrate in biological environment with preserving the quality factor in high value. As demonstrated FEM results this sensor brings two detectable parameters in liquid medium first of all, resonant frequency and secondly average displacement. Because this sensor is based on CMUT technology can be used in large arrays for multi analytes which is other benefit of described sensor.

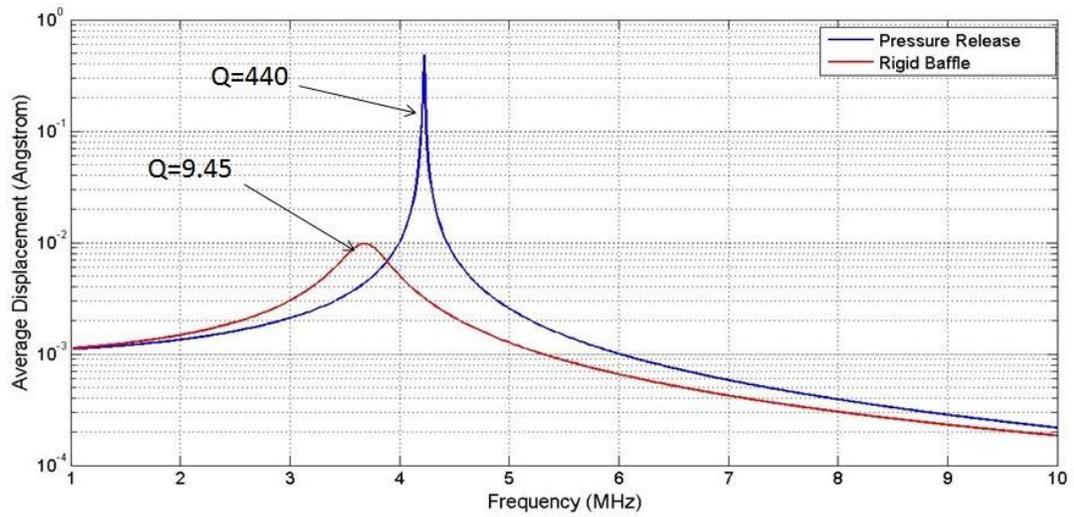

**Figure 38:** Average membrane displacement as a function frequency for two different boundary conditions on the baffle.

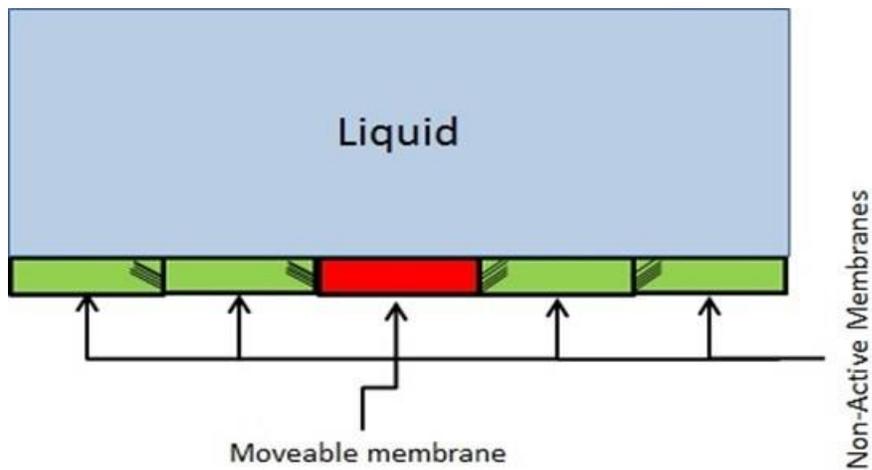

**Figure 39:** 40 micro-meter passive membranes for providing the pressure release condition.

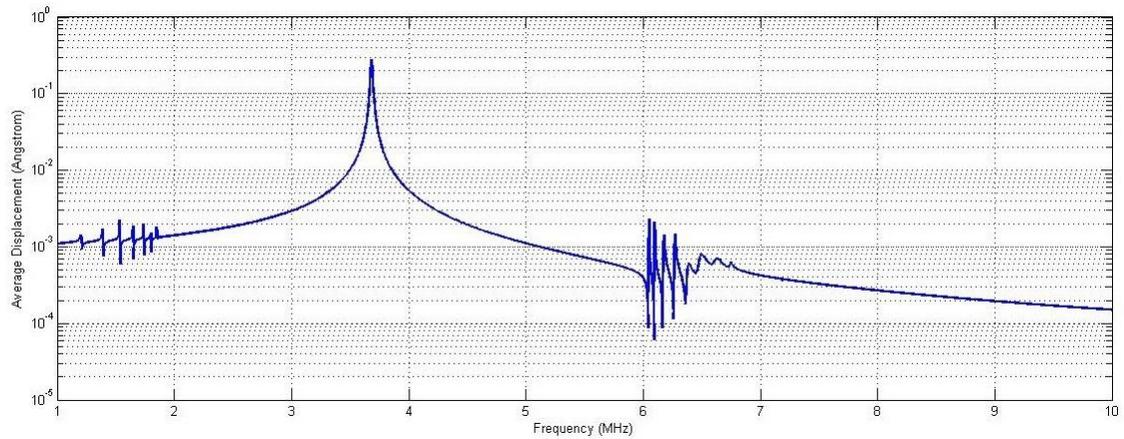

**Figure 40:** Average displacement of the membrane by providing the pressure release boundary condition. Thus, several non-active membranes added to beside the active membrane.

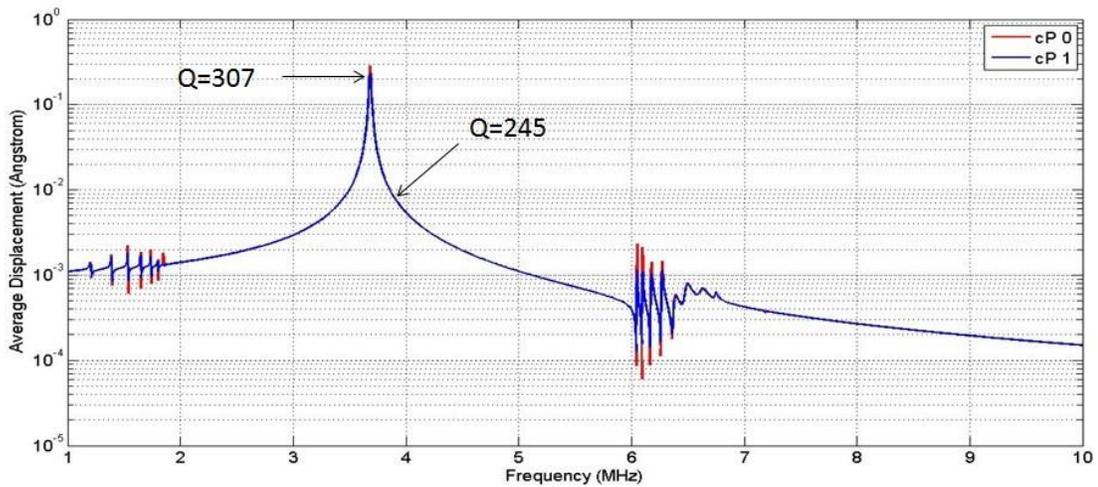

**Figure 41:** Average displacement of the membrane when pressure release condition provided. 1 centi-poise viscosity added to water properties and 20% reduction in quality factor for proposed sensor observed.

# CHAPTER VI

# CONCLUSION AND FUTURE WORK

In first part of this master thesis, we have demonstrated blood coagulation time measurement using very high frequency ultrasound in disposable low cost cartridges. The explained method requires only 1 $\mu$L of whole blood and it is appropriate for patient self-testing. This amount of blood sample can be drawn by a lancet from fingers head simply. The cartridge cost is relatively low since it uses low cost materials (glasses) and requires simple MEMS fabrication techniques. To further decrease the cost of the cartridges, the transducer can be fabricated on a separate substrate and the substrate can be brought into contact with the cartridge in a reader unit. Because this approach gets rid of both the transducer and the electrical contacts, it greatly reduces the cost of the cartridge.

In present study, all results showed, that the ultrasonic system is capable of monitoring whole blood reaction with coagulation activating and inactivating agent, thus it is useful for evaluating blood hypercoagulability and hypocoagulability which other coagulation test operates just based on hypocoagulability. Furthermore, it's simple and fast procedure for measuring aggregation of platelets by a drop sample.

It has been demonstrated that ultrasound transducers can be used to detect the viscosity (and velocity) changes that happen during the blood coagulation. The attenuation changes in blood coagulation time and proportional that the voltage amplitude of reflection changes. Therefore, monitor the blood coagulation actual time by using this Bio-MEMS, which is easily applicable.

Measurements using remotely coupled transducers have been demonstrated in earlier study [46] at lower frequencies. However, the extension of the method for high frequencies is not obvious and requires further research. Another advantage of the described system is the fact that one can perform multiple measurements from the same blood sample to increase the reliability of the test using multiple transducers. In other words, the explained system can be operated for multiple detections in same time. Thus, multiple transducers can be connected to the RF electronics through a switch without increasing the complexity of the reader unit. Furthermore, cartridges can be configured to allow transducers to measure blood in different channels as a array. By doing that, one can perform various coagulation tests with different reagents on the same cartridge using the same blood sample.

Previously, ultrasound had been used to perform mixing in the microfluidic channels. As described in the earlier research [47, 26] one can use radiation pressure to mix the blood with dried reagents in the channels. This should increase the repeatability and reliability of the tests since the red blood cell sedimentation will be prevented as well as reagents will be mixed with blood uniformly.

Future work will include manufacturing several micro-fluid channels which are integrated with high ultrasonic transducers. This method can be dried reagents to perform PT and aPTT tests. To increase the reliability and repeatability of the tests based on described system in this master thesis one can use the transducers to mix the blood sample with these agents. Same transducer can be used to perform both mixing and measurement as outlined in the previous works. In other words, if mixing and measurements provided on the chip we can call it lab on a chip because of multi medical function can be measured by a small and portable system.

In second part of this master thesis, new resonant mass sensor based on CMUTs for operating in liquid for biological detection introduced. Since in biological environment the acoustic impedance is high then high value of acoustic losses in liquids medium effect on quality factor. The described method can overcome low sensitivity of resonant mass sensors problem in liquids environment. This method used pressure release condition by using several passive membranes beside the sensing element (Active membrane) and it could provide the high quality resonant mass sensor for immersion in liquids. The main advantage of the suggested structure for resonant mass sensors based on CMUT is just one cell vibrates as a sensing element which is reduced the complexity of electronic connection. Other advantage proposed sensor is it can be manufactured by simple MEMS technology as large arrays for multiple measurements at the same time. So, reliability and repeatability will be increased by this procedure.

After fabrication steps are finished the sensor will be tested under water. To measure the membrane displacement a laser vibrometer will be used. The membrane will be vibrated by applying AC voltage on top of a DC bias voltage. Therefore an optical detection is required. The integrated optical detection will be implemented similar to the method in reference [48]. However, this is not single method for measuring average membrane displacement. Several techniques such as capacitive, magnetic and piezoelectric are available for this purpose. The membranes will be built on a quartz substrate which is transparent. Bottom electrode will form a grating structure. A laser diode will be aligned to the membrane from the bottom side. The reflected light from the grating and the membrane will interfere creating diffraction orders. By measuring these orders one can monitor the membrane displacement. Many membranes can be built on a substrate as shown in figure 42. This will be future work of this thesis and this platform could be the basis of many biological applications.

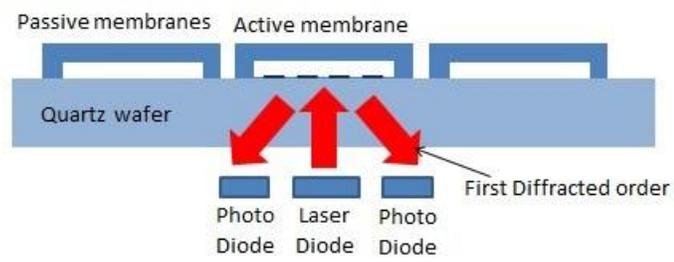

**Figure 42:** Sensor platform with integrated optical read out.

# VITA

He was born in 1991 in Hamedan, Iran and graduated with bachelor of Biomedical Engineering (Bio-Electric) in the faculty of Electrical and Electronics Engineering in Sahand University of Technology Tabriz, Iran with first rank among biomedical engineering students in 2013.

He pursued his M.Sc. degree in Electrical and Electronics Engineering in Ozyegin University, Istanbul, Turkey under supervision of Prof. Goksenin Yaralioglu, where he focuses on design and development of a microfluidic channels that integrated with the ultrasonic transducer for sensing the blood properties and design a impedance matching circuit for various high frequencies range.

Furthermore, he could attend to 14th international MUT conference (Deresden, Germany, 19th May 2015) and presented an oral paper from his thesis results. Also, he could submit a research paper to a journal from his master thesis. His research interests are: biosensors, Bio-MEMS and ultrasonic applications.

In addition, parallel to his research he found chance to be teaching assistant for Analog Electronics and Circuit Analysis courses in Ozyeging University for two years.

He will pursue his PhD degree in Electrical and Computer Engineering field in Memorial University, Newfoundland, Canada.